\documentclass[11pt]{article}
\usepackage{geometry}	
\usepackage{graphicx} 
\usepackage{epstopdf}
\usepackage{url}	
\usepackage{setspace}\spacing{1.213}	
\usepackage[bottom]{footmisc}		
\usepackage{rotating,float}		
\usepackage[justification=centering,font=small]{caption} 
\usepackage{verbatim}  
\usepackage{adjustbox} 
\usepackage{relsize}   
\usepackage{amsmath,amssymb}\allowdisplaybreaks[1] 

\usepackage{appendix} 
\usepackage{placeins}
\usepackage[small,compact]{titlesec}  
\usepackage[bookmarks=true]{hyperref} 
\usepackage{bookmark}
\usepackage{pdflscape}
\usepackage{url}
\usepackage{ragged2e}
\usepackage{bm}
\usepackage{caption}
\usepackage{subcaption}

\usepackage{xcolor}	
\hypersetup{
	colorlinks,
	linkcolor={red!50!black},
	citecolor={blue!50!black},
	urlcolor={blue!80!black}
}

\usepackage[authoryear,round]{natbib}

\usepackage{amsthm}
\theoremstyle{plain}

\theoremstyle{definition}

\newtheoremstyle{indenteddefinition}
	{}
	{}
	{\hangindent=2em}
	{}
	{\bfseries}
	{.}
	{.5em}
	{}
\theoremstyle{indenteddefinition}
\newtheorem{assump}{}
\newcounter{assumptiongroup}\stepcounter{assumptiongroup}
\renewcommand{\theassump}{\Alph{assumptiongroup}.\arabic{assump}}

\title{Counting Defiers: Examples from Health Care\thanks{Previous versions of this paper have been circulated under different working paper numbers and different titles including ``Counting Defiers,''  ``A Model of a Randomized Experiment with an Application to the PROWESS
Clinical Trial,'' and ``General Finite Sample Inference for Experiments with Examples from Health Care'' \citep{kowalski2019a,kowalski2019b}. I thank Neil Christy, Tory Do, Simon Essig Aberg, Bailey Flanigan, Pauline Mourot, Srajal Nayak, Sukanya Sravasti, and Matthew Tauzer for excellent research assistance. Don Andrews, Susan Athey, Victoria Baranov, Steve Berry, Michael Boskin, Kate Bundorf, Xiaohong Chen, Victor Chernozhukov, Peng Ding, Brad Efron, Ivan Fernandez-Val, Michael Gechter, Matthew Gentzkow, Florian Gunsilius, Andreas Hagemann, Jerry Hausman, Han Hong, Guido Imbens, Daniel Kessler, Jonathan Kolstad, Ang Li, Aprajit Mahajan, Charles Manski, Elena Pastorino, John Pepper,  Demian Pouzo, Jann Spiess, Edward Vytlacil, Stefan Wager, Christopher Walters, David Wilson, and seminar participants at the Advances with Fields Experiments Conference at the University of Chicago, the AEA meetings, the Essen Health Conference, Notre Dame, the Stanford Hoover Institution, UCLA, UVA, the University of Zurich, the Yale Cowles Summer Structural Microeconomics Conference, and the Y-RISE Evidence Aggregation and External Validity Conference provided helpful comments.  I thank Charles Antonelli,  Bennett Fauber, Corey Powell, and Advanced Research Computing at the University of Michigan, as well as Misha Guy, Andrew Sherman, and the Yale University Faculty of Arts and Sciences High Performance Computing Center. I also thank my parents and sister for their support.}} 
\author{A. E. Kowalski}

\begin{document}
	\maketitle
	\singlespacing
	\begin{center}

\end{center}

\begin{abstract}
\noindent  I propose a finite sample inference procedure that uses a likelihood function derived from the randomization process within an experiment to conduct inference on various quantities that capture heterogeneous intervention effects.  One such quantity is the number of defiers---individuals whose treatment runs counter to the intervention. Results from the literature make informative inference on this quantity seem impossible, but they rely on different assumptions and data.  I only require data on the cross-tabulations of a binary intervention and a binary treatment.  Replacing the treatment variable with a more general outcome variable, I can perform inference on important quantities analogous to the number of defiers. I apply the procedure to test safety and efficacy in hypothetical drug trials for which the point estimate of the average intervention effect implies that at least 40 out of 100 individuals would be saved.  In one trial, I infer with 95\% confidence that at least 3 individuals would be killed, which could stop the drug from being approved. 
\end{abstract}

\pagebreak
\section{Introduction}
\onehalfspacing

Suppose a clinical trial shows that an intervention increases survival. Using standard inference procedures, we conclude that the intervention must save the lives of some individuals in the trial. But we care about safety as well as efficacy.  Can we also test the hypothesis that no individuals in the trial would be killed if they were randomized into the intervention arm? I propose a finite sample inference procedure that allows us to do so.

My main contribution is that I can conduct inference on the number of ``defiers,'' individuals whose treatment runs counter to the intervention \citep{balke1993,angrist1996}, and analogous quantities. The inference that I conduct on the number of defiers is informative in the sense that it can reject the null hypothesis that there are zero defiers even when the point estimate of the average intervention effect indicates that there are more compliers than defiers.  I am not aware of any other procedure that can do so using the same data.   By performing inference on the number of defiers and analogous quantities, I offer a novel test of the LATE monotonicity assumption of \citet{imbens1994} and the related monotone response assumption of \citet{manski1997monotone}.

However, my contribution is not limited to informative inference on the number of defiers and analogous quantities.  The inference procedure allows me to provide p-values and confidence intervals on various quantities that capture heterogeneous intervention effects.  For some of these quantities, I am not aware of any previous inference procedures that are informative.  For others, previous inference procedures are informative but approximate or only applicable to a single quantity.  I can use the same procedure that I use to perform inference on the number of defiers to perform inference on the average intervention effect considered by the  \citet{neyman1923} null hypothesis  and the fraction affected by the intervention in either direction considered by the \citet{fisher1935} null hypothesis.  I can also test multiple hypotheses simultaneously.   

The inference procedure that I propose is a ``finite sample'' inference procedure because it focuses on the finite sample of individuals in an experiment.  In their early work on experiments, \citet{neyman1923} and  \citet{fisher1935} were both interested in exact calculations on finite samples, but they could only perform them in very small samples, so the field of statistics largely adopted simplifying asymptotic assumptions.  Such assumptions yield approximations that obscure information, sometimes precluding informative inference.  Advances in computational power allow me to forgo simplifying asymptotic assumptions and perform exact calculations, even in large samples. 

For data, the procedure that I propose only requires the cross-tabulations of a binary intervention and a binary outcome.  To perform inference on the number of defiers---individuals whose treatment runs counter to the intervention---I interpret the outcome variable as takeup of a treatment.  The required data consist of four numbers: the number of treated and untreated individuals in the intervention and control arms.  I refer to the four numbers as the ``data configuration.''  

I specify how the data configuration is generated with a model of a randomized experiment that has two assumptions, where the first assumption draws on the concept of potential outcomes \citep{neyman1923, rubin1974, rubin1977, holland1986}.  It requires that each individual has one potential outcome in intervention and one potential outcome in control. \citet{cox1958} refers to this assumption as the ``no interference'' assumption, and \citet{rubin1980} refers to it as the ``stable unit treatment value assumption'' (SUTVA). For a binary outcome, this assumption implies that there are four types of individuals, corresponding to four ``principal strata'' of \citet{frangakis2002}.  When the outcome represents takeup of a treatment, one of the types represents defiers.  Defiers have a potential outcome of zero in intervention because they are untreated and a potential outcome of one in control because they are treated, so the intervention effect for defiers is negative one. In the terminology of \citet{angrist1996}, the other three potential outcome types represent ``compliers''  whose treatment aligns with the intervention (intervention effect of positive one), ``always takers'' who are treated regardless of the intervention (intervention effect of zero), and ``never takers'' who are untreated regardless of the intervention (intervention effect of zero). I refer to the vector of the number of individuals of each type as the ``potential outcome type configuration.''   The various quantities on which I perform inference are functions of the potential outcome type configuration, and they therefore capture heterogeneous intervention effects.

The second assumption specifies the randomization process within the experiment.  This randomization process is a feature of the experimental design.  There are many different possible randomization processes.  For example, the experimenter could flip a coin for each individual.  Alternatively, the experimenter could draw balls out of an urn. Because the experimenter can control the randomization process, the experimenter can also build a compelling justification for the assumption that follows from it.  

As the cornerstone for inference, I use the two assumptions of the model to derive the likelihood function of the potential outcome type configuration given the data configuration.  I begin by deriving a general likelihood function that arises from any  known randomization process.   I then present two specific likelihood functions that arise from two specific randomization processes. I show that the second function is equivalent to a likelihood function that appears in  \citet{copas1973}.    

I use the likelihood function for hypothesis testing and construction of confidence intervals rather than estimation because there can be more than one potential outcome type configuration that maximizes the likelihood of the data configuration.  For example, consider an experiment with 100 individuals.  There are 176,851 possible ways to divide 100 individuals into four groups.  Therefore, there are 176,851 possible data configurations and 176,851 possible potential outcome type configurations, yielding $176,851^2$---over 31 billion---possible combinations.  However, some potential outcome type configurations cannot give rise to some data configurations. Among those that can give rise to a given data configuration, the specific randomization process dictates that some are more likely than others.  By evaluating a specific likelihood function over 31 billion times, I know that the maximum likelihood estimate of the potential outcome type configuration is multi-valued for 14,940 of the 176,851 possible data configurations.

To conduct inference, I evaluate a specific likelihood function for every combination of the data and potential outcome type configurations to construct a likelihood ratio test statistic and its exact distribution under the null hypothesis.  I conduct exact inference in the sense that I control the type I error rate, the probability of a false rejection, within the finite sample, even when I test multiple hypotheses simultaneously.  I also construct confidence intervals.

I demonstrate that the inference procedure that I propose can conduct informative inference on the number of defiers by applying it to hypothetical drug trials. I show that the number of individuals who would be killed by the intervention in a trial is analogous to the number of defiers.  I present data from two trials with 100 individuals. Each trial has a different data configuration.  However, both trials yield the same point estimate of the average intervention effect, which indicates that there are 40 more individuals who would be saved than killed.  This point estimate is highly statistically significant in both trials.  In one trial, I cannot reject the null hypothesis that no one would be killed at any level. In the other, I reject the same null hypothesis at the 2.8\% level, and I construct a 95\% confidence interval that shows that at least three individuals would be killed.  In the same trial, in a joint test of safety and efficacy, I reject the composite null hypothesis that some individuals would be saved and no individuals would be killed at the 7.6\% level.  These results demonstrate that the ability to conduct informative inference on defiers and related quantities could change the drug approval process.

Several results make inference on the number of defiers seem impossible unless the point estimate of the average intervention effect indicates that there are more defiers than compliers, and I engage with them to investigate how informative inference on the number of defiers is possible.  I begin with the Boole-Fr\'echet-Hoeffding  bounds \citep{boole1854, frechet1957, hoeffding1940, hoeffding1941}.  These bounds have been applied numerous times in the treatment effects literature \citep{heckman1997, manski1997mixing, tian2000, zhang2003, fan2010, mullahy2018, ding2019}.  I review using my notation that the point estimate of the Boole-Fr\'echet-Hoeffding lower bound on the share of defiers is zero when the point estimate of the average intervention effect indicates there are more compliers than defiers. Inference using these bounds cannot reject the null hypothesis of zero defiers in this case, so it is not informative.   \citet{ruschendorf1981} shows that these  bounds are the tightest possible given the data and assumptions, making informative inference on the number of defiers seem impossible. 

I perform informative inference on the number of defiers by requiring different assumptions and more data than applications of the Boole-Fr\'echet-Hoeffding bounds. For assumptions, though I require that the randomization process is known, the Boole-Fr\'echet-Hoeffding bounds can be applied under a weaker assumption that requires the intervention is independent from the potential outcomes (see, for example, \citet{tian2000}).  For data, I require four numbers, whereas applications of the Boole-Fr\'echet-Hoeffding bounds require only two.  To understand the consequence of requiring more data, I derive an alternative likelihood function given my assumptions and the limited data required by the Boole-Fr\'echet-Hoeffding bounds, and I use it for inference on hypothetical clinical trial data.  The results that I obtain are still informative about the number of defiers, but they are less precise than results that use the full data, so the limited data cannot fully explain why informative inference is impossible.  Therefore, my assumptions must play an important role in my relative ability to perform informative inference on the number of defiers.

My assumptions also appear to play an important role in differentiating the procedure I propose from other impossibility results that require more data than I require.  Specifically, \citet{balke1997} and \citet{heckman2001instrumental} require data on an intervention, an outcome, \emph{and} a treatment that is empirically distinct from the outcome to obtain bounds on the average treatment effect, and they show that their bounds are the tightest possible under their assumptions.  \citet{kitagawa2015} builds on testable restrictions implied by their bounds and related testable restrictions of \citet{imbens1997} and \citet{heckman2005} to test the LATE monotonicity assumption jointly with other assumptions.  He obtains the ``strongest testable implication for instrument validity,'' making informative inference on the number of defiers seem impossible.  Other joint tests of the LATE monotonicity assumption also require more data  \citep{richardson2010,huber2013,huber2015testing,mourifie2017,machado2019}.  The required assumptions vary, but to my knowledge, applications of all previous tests joint tests of LATE monotonicity impose a simplifying asymptotic assumption.
   
I demonstrate that the lack of a simplifying asymptotic assumption is integral to my ability to perform informative inference on the number of defiers.  I do so by adding an explicit asymptotic assumption to my model and showing that it is sufficient to yield an impossibility result.   I begin by deriving an alternative likelihood under the model plus the asymptotic assumption.  As part of the derivation, I demonstrate that the resulting likelihood is a weighted average of the general likelihood that I obtain under the model alone, thereby demonstrating that the asymptotic assumption is a simplifying assumption that obscures information.  Next, I demonstrate that under two different specific randomization processes, the resulting likelihoods are equivalent to likelihoods derived by \citet{barnard1947} and \citet{kline2020}. While these likelihoods vary with the average intervention effect and have been used for inference on that quantity, I show that they cannot be used for informative inference on the number of defiers within a sample of \emph{any size}. Asymptotic inference simplifies inference on the average intervention effect while precluding informative inference on the number of defiers at all sample sizes. There is no sense in which finite sample inference on defiers is worse in large samples than it is in small samples. The important distinction is not between large and small samples, but between asymptotic and finite sample inference.     

The previous finite sample inference literature does not impose simplifying asymptotic assumptions, but to my knowledge, it also does not perform informative inference on the number of defiers.  This literature began with the \citet{fisher1935} exact test.  Through approximations of Fisher's exact test \citep{dwass1957}, the subsequent ``randomization inference'' literature has developed inference methods based on the randomization process (see \citet{canay2017} for a recent example). However, this literature treats defiers and analogous quantities as nuisance parameters with respect to inference on other quantities \citep{copas1973, chung2013,chiba2015,rigdon2015,ding2016,li2016,ding2019,wu2020}.  My primary contribution to this literature is that I aim to perform inference on the number of defiers, and I recognize that it is possible to do so by deriving a likelihood for use with a likelihood ratio test statistic.  Other test statistics, such as the point estimate of the average intervention effect, do not explicitly depend on the likelihood that I derive. My secondary contribution to the finite sample inference literature is that I can use the same procedure to conduct inference on quantities other than defiers.  

The rest of this paper proceeds as follows. Section~\ref{sec:model} introduces the model and derives the likelihood. Section~\ref{sec:hypothesis} describes the hypothesis testing procedure and the construction of confidence intervals. Section~\ref{sec:fda} presents an application to hypothetical drug trials.  Section~\ref{sec:lit} discusses how the proposed inference procedure relates to impossibility results from the literature.  Section~\ref{sec:conclusion} concludes.	

\section{Model: Derivation of Likelihood} \label{sec:model} \label{sec:derivation}

Consider a randomized experiment with sample size $s$, where $s$ is a positive integer. There is a binary intervention $Z$ and a binary outcome $D$ that represents treatment. Each individual is assigned to either intervention ($Z=1$) or control ($Z=0$), and each individual is either treated ($D=1$) or untreated ($D=0$).

\begin{figure}[!hbt]
	\caption{Matrix that Relates Elements of the Potential Outcome Type Configuration \\to Elements of the Data Configuration When the Outcome Represents Treatment}
	\centering
	\includegraphics[width=1\linewidth]{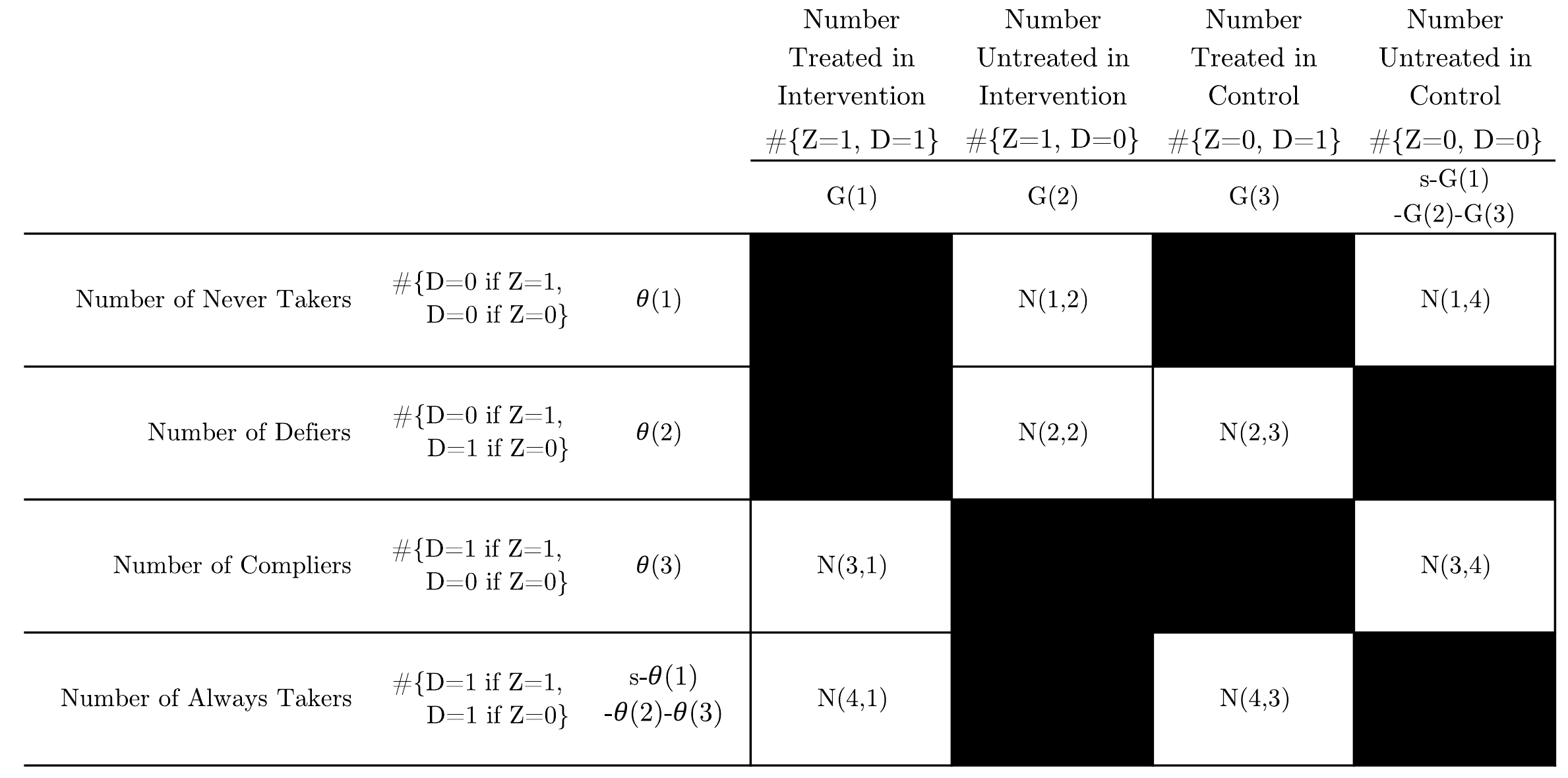}
	\label{fig:matrix} \\
	\vspace{-10pt}
	\begin{minipage}{1\linewidth}
	\scriptsize
	\emph{Note}. $s$ represents the sample size. The potential outcome type configuration $\bm{\theta}$ is the vector of $(\theta(1), \theta(2), \theta(3), s-\theta(1)-\theta(2)-\theta(3))$. The data configuration $\bm{G}$ is the vector of $(G(1),G(2),G(3),s-G(1)-G(2)-G(3))$. $D$ represents treatment, and $Z$ represents assignment to the intervention arm. $\#\{\cdot\}$ denotes the number of elements in the set $\{\cdot\}$. $N(j,k)$ denotes the number of individuals in element $j$ of the potential outcome type configuration and element $k$ of the data configuration. All shaded cells equal zero.
	\end{minipage}
\end{figure}

I impose the following assumption, which represents the ``no interference'' assumption \citep{cox1958} or the ``stable unit treatment value assumption'' (SUTVA) \citep{rubin1980}:
\begin{assump} \label{sutva}(No interference). Each individual has one potential outcome in the control arm and one potential outcome in the intervention arm.
\end{assump} 
\noindent Under \ref{sutva}, each individual has a potential outcome type, where each potential outcome type is characterized by a combination of a potential outcome in the control arm and a potential outcome in the intervention arm.  The matrix in Figure~\ref{fig:matrix} includes a separate row for each of the four possible potential outcome types.  The first row gives the number of never takers ($D=0$ if $Z=1$ and $D=0$ if $Z=0$), denoted $\theta(1)$; the second row gives the number of defiers ($D=0$ if $Z=1$ and $D=1$ if $Z=0$), denoted $\theta(2)$; the third row gives the number of compliers ($D=1$ if $Z=1$ and $D=0$ if $Z=0$), denoted $\theta(3)$; and the fourth row gives the number of always takers ($D=1$ if $Z=1$ and $D=1$ if $Z=0$), denoted $s-\theta(1)-\theta(2)-\theta(3)$.  The potential outcome type configuration consists of the four nonnegative integers $\theta(1)$, $\theta(2)$, $\theta(3)$, and $s-\theta(1)-\theta(2)-\theta(3)$, which I represent with $\bm{\theta}$, in bold to indicate that the potential outcome type configuration is a vector. Although the potential outcome type configuration consists of four numbers, because the sample size $s$ is known, the potential outcome type configuration has only three unknown elements: $\theta(1), \theta(2),$ and $\theta(3)$.  

The matrix in Figure~\ref{fig:matrix} includes a separate column for each element of the data configuration.  The first column includes the number of individuals observed treated in the intervention arm ($Z=1$ and $D=1$), denoted $G(1)$; the second column includes the number of individuals observed untreated in the intervention arm ($Z=1$ and $D=0$), denoted $G(2)$; the third column includes the number of individuals observed treated in the control arm ($Z=0$ and $D=1$), denoted $G(3)$; and the fourth column includes the number of individuals observed untreated in the control arm ($Z=0$ and $D=0$), denoted $s-G(1)-G(2)-G(3)$. I denote $G(1)$, $G(2)$, and $G(3)$ with capital letters because they are random variables. The data configuration consists of the four nonnegative integers $G(1)$, $G(2)$, $G(3)$, and $s-G(1)-G(2)-G(3)$, which I represent with $\bm{G}$, in bold to indicate that the data configuration is a vector. Although the data configuration consists of four numbers, conditional on the sample size $s$, it can be fully represented by the three elements $G(1)$, $G(2)$, and $G(3)$. In what follows, I denote a realization of $G(\cdot)$ with $g(\cdot)$ and a realization of $\bm{G}$ with $\bm{g}$. 

The matrix in Figure~\ref{fig:matrix} relates the elements of the potential outcome type configuration to the elements of the data configuration.  $N(j,k)$ is the nonnegative integer that represents the number of individuals in element $j$ of the potential outcome type configuration and element $k$ of the data configuration. The matrix of all $N(j,k)$ has 16 total cells.  However, the eight shaded cells cannot have any individuals in them.  For example, consider the upper left cell.  Never takers, potential outcome type $j=1$, would be untreated regardless of assignment to the intervention arm, and individuals in element $k=1$ of the data configuration are observed treated in the intervention arm.  By definition, never takers will not be observed treated in the intervention arm, so it must be the case that $N(1,1)=0$. Similarly, $N(1,3)=0$, because never takers $j=1$ will not be observed treated in the control arm $k=3$.  The logic for the other shaded cells proceeds similarly. 

The purpose of the model is to allow me to derive an expression for the likelihood of the potential outcome type configuration $\bm{\theta}$ conditional on the data configuration $\bm{g}$, the sample size $s$, and any other parameters that govern the randomization process from the sample into the intervention $\bm{\gamma}$.  In the equations below, I begin by expressing the likelihood in terms of the probability of the data configuration.  Then, under the assumption of no interference (\ref{sutva}), I move from (\ref{eq:start}) to (\ref{eq:eight}) by expressing the elements of the data configuration $G(1)$, $G(2)$, and $G(3)$ in terms of the cells in the corresponding columns in Figure~\ref{fig:matrix}:
\begin{align}
\mathcal{L}(\bm{\theta}\mid \bm{g},s,\bm{\gamma}) 
&= P\big(\bm{G}=\bm{g} \mid \bm{\theta},s,\bm{\gamma}\big) \nonumber\\
&= P\big(G(1)=g(1),G(2)=g(2),G(3)=g(3)\mid \bm{\theta},s,\bm{\gamma}\big) \label{eq:start} \displaybreak[2]\\
&= P\bigg(N(3,1)+N(4,1) = g(1), \nonumber \\*
&\quad\qquad N(1,2)+N(2,2) = g(2), \nonumber \\*
&\quad\qquad N(2,3)+N(4,3) = g(3)\mid \bm{\theta},s,\bm{\gamma}\bigg) \label{eq:eight} 
\end{align}

\noindent I can simplify the expression	 by recognizing that, within an element $j$ of the potential outcome type configuration $\bm{\theta}$, the number of individuals in the control arm equals the total number of individuals minus the number of individuals in the intervention arm. Therefore, in (\ref{eq:subs}), I express (\ref{eq:eight}) in terms of the four random variables $N(1,2)$, $N(2,2)$, $N(3,1)$ and $N(4,1)$ that represent the number of individuals assigned to the intervention arm of each of the four potential outcome types.  I simplify the expression by rearranging the terms of (\ref{eq:subs}) so that the random variables $N(1,2)$, $N(2,2)$, $N(3,1)$, and $N(4,1)$ are on the left side of each equation, which I express in (\ref{eq:four}):
\begin{align}
\mathcal{L}(\bm{\theta}\mid \bm{g},s,\bm{\gamma}) 
&= P\bigg(N(3,1)+N(4,1) = g(1), \nonumber \\*
&\quad\qquad N(1,2)+N(2,2) = g(2), \nonumber \\*
&\quad\qquad \theta(2)-N(2,2)+s-\theta(1)-\theta(2)-\theta(3)-N(4,1) = g(3) \mid \bm{\theta}, s,\bm{\gamma}\bigg) \label{eq:subs}\\
&= P\bigg(N(3,1)+N(4,1) = g(1), \nonumber \\*
&\quad\qquad N(1,2)+N(2,2) = g(2), \nonumber \\*
&\quad\qquad N(2,2)+N(4,1) = s-\theta(1)-\theta(3)-g(3) \mid \bm{\theta},s,\bm{\gamma}\bigg). \label{eq:four}
\end{align}

\noindent For a given potential outcome type configuration $\bm{\theta}$, there can be multiple realizations of the random variables $N(1,2)$, $N(2,2)$, $N(3,1)$ and $N(4,1)$ consistent with the data configuration $\bm{g}$.  The probability expressed in (\ref{eq:four}) can be written as the sum of the probabilities of each of these realizations.  I express that sum in (\ref{eq:ell}), indexing each realization by setting $N(1,2)=\ell$ for values of $\ell$ ranging from zero to $\theta(1)$.  Then, in (\ref{eq:ten}), I rearrange (\ref{eq:ell}) to express it in terms of the joint distribution of the number of individuals assigned to the intervention arm within each of the four potential outcome types $N(1,2)$, $N(2,2)$, $N(3,1)$, and $N(4,1)$:
\begin{align}
\mathcal{L}(\bm{\theta}\mid \bm{g},s,\bm{\gamma})
&= \sum_{\ell=0}^{\theta(1)} P\bigg(N(1,2) = \ell, \nonumber\\*
&\qquad\qquad\;\ N(3,1) + N(4,1) = g(1), \nonumber\\*
&\qquad\qquad\;\ N(1,2) + N(2,2) = g(2), \nonumber\\*
&\qquad\qquad\;\ N(2,2) + N(4,1) = s - \theta(1) - \theta(3) - g(3) \mid \bm{\theta}, s, \bm{\gamma} \bigg) \label{eq:ell}\\
&= \sum_{\ell=0}^{\theta(1)} P\bigg(N(1,2)=\ell, \nonumber \\*
&\qquad\qquad\;\ N(2,2) = g(2)-\ell, \nonumber \\*
&\qquad\qquad\;\ N(3,1) = \theta(1)+\theta(3)+g(1)+g(2)+g(3)-s-\ell, \nonumber\\*
&\qquad\qquad\;\ N(4,1) =  s+\ell-\theta(1)-\theta(3)-g(2)-g(3) \mid \bm{\theta},s,\bm{\gamma}\bigg). \label{eq:ten}
\end{align}

\noindent To complete the derivation of the likelihood, I impose a second assumption, which specifies the randomization process within the experiment:
\begin{assump} \label{rand}(Known Randomization Process within the Experiment: General Case). Individuals in the
intervention arm are selected from the sample through a known process, which yields a known function $f$ specifying the joint probability mass function of $N(1,2)$, $N(2,2)$, $N(3,1)$, and $N(4,1)$ conditional on the potential outcome type configuration $\bm{\theta}$, the sample size $s$, and other parameters that govern the randomization process from the sample into the intervention $\bm{\gamma}$:
\begin{align*}
&f\big(n(1,2),n(2,2),n(3,1),n(4,1)\mid \bm{\theta},s,\bm{\gamma}\big) \\*
&\qquad \equiv P\bigg(N(1,2)=n(1,2), N(2,2)=n(2,2), N(3,1)=n(3,1), N(4,1)=n(4,1)\mid \bm{\theta}, s,\bm{\gamma}\bigg).
\end{align*}
\end{assump}

\noindent Under \ref{rand}, I can substitute $f$ for the joint distribution of the numbers of each potential outcome type randomized into intervention in (\ref{eq:ten}) to obtain a known functional form for the likelihood:
\begin{align}
\mathcal{L}(\bm{\theta}\mid \bm{g},s,\bm{\gamma})
&= \sum_{\ell=0}^{\theta(1)} f\big(\ell,g(2)-\ell,\theta(1)+\theta(3)+g(1)+g(2)+g(3)-s-\ell, \nonumber\\*
&\qquad\qquad s+\ell-\theta(1)-\theta(3)-g(2)-g(3) \mid \bm{\theta},s,\bm{\gamma}\big). \label{eq:f}
\end{align}

\noindent Equation (\ref{eq:f}) provides a general expression for the likelihood of the potential outcome type configuration $\bm{\theta}$.  The specific functional form of the likelihood depends on the known randomization process. There are many different possible randomization processes. I present two cases here, but these are not meant to be exhaustive. For the first case, suppose the randomization process from the sample into the intervention arm is such that:

\renewcommand{\theassump}{A.2(iid)}
\begin{assump} \label{iid}(Known Randomization Process within the Experiment: IID Case). Individuals in the intervention arm are selected from the sample through the flip of a weighted coin such that $Z$ is independently and identically distributed (IID), the intended fraction in intervention is $p\in (0,1)$, and the functional form of $f$ is the product of four independent binomial distributions parameterized by the potential outcome type configuration $\bm{\theta}$, the sample size $s$, and the intended fraction randomized into intervention $p$:
\begin{align*}
f\big(n(1,2),n(2,2),n(3,1),n(4,1) \mid \bm{\theta},s,p\big) 
&= \mathrm{binom}\big(n(1,2),\theta(1),p\big)  \mathrm{binom}\big(n(2,2),\theta(2),p\big)\\*
&\quad\times \mathrm{binom}\big(n(3,1),\theta(3),p\big) \\*
&\quad\times \mathrm{binom}\big(n(4,1),s-\theta(1)-\theta(2)-\theta(3),p\big),
\end{align*}
where  $\mathrm{binom}(a,b,p)$ is the binomial probability mass function for a nonnegative integer number of successes $a$ among a nonnegative integer number of trials $b\geq a$ and probability of success $p\in(0,1)$:
\begin{align*}
\mathrm{binom}(a,b,p) = P(A=a\mid b,p) = \binom{b}{a} p^a (1-p)^{b-a}.
\end{align*} 

\end{assump} 

\noindent As one algorithm to implement this randomization process, the experimenter could assign each individual in the experiment a uniform random number from 0 to 1 and choose individuals with a number less than the intended fraction randomized into intervention $p$ to be assigned to the intervention arm. To complete the derivation of the specific likelihood for the IID case, I substitute the functional form of $f$ given by \ref{iid} into the general likelihood expression in (\ref{eq:f}). Therefore, under \ref{sutva} and \ref{iid}, the likelihood of the potential outcome type configuration $\bm{\theta}$ given the data configuration $\bm{g}$, the sample size $s$, and the intended fraction randomized into intervention $p$ is as follows:
\begin{align}
\mathcal{L}(\bm{\theta} | \bm{g},s,p)&= \sum_{\ell=0}^{\theta(1)}\mathrm{binom}\big(\ell,\theta(1),p\big) \nonumber \\*
&\qquad \times \mathrm{binom}\big(g(2)-\ell,\theta(2),p\big) \nonumber \\*
&\qquad \times \mathrm{binom}\big(\theta(1)+\theta(3)+g(1)+g(2)+g(3)-s-\ell,\theta(3),p\big) \nonumber \\*
&\qquad \times \mathrm{binom}\big(s+\ell-\theta(1)-\theta(3)-g(2)-g(3),s-\theta(1)-\theta(2)-\theta(3),p\big). 
\end{align}

For the second case, suppose the randomization process from the sample into the intervention arm is such that:
\renewcommand{\theassump}{A.2(urn)}
\begin{assump} \label{urn}(Known Randomization Process within the Experiment: Urn Case). Individuals in the intervention arm are selected from the sample by drawing $m$ names from an urn, where $m$ is a positive integer less than the sample size $s$.  This process implies that $G(1)+G(2)$ is constrained to equal $m$, and the functional form of $f$ is a multivariate hypergeometric distribution parameterized by the potential outcome type configuration $\bm{\theta}$, the sample size $s$, and the number of individuals randomized into the intervention arm $m$:
\begin{align*}
f\big(n(1,2),n(2,2),n(3,1),n(4,1) \mid \bm{\theta},s,m) 
&= \binom{\theta(1)}{n(1,2)} \binom{\theta(2)}{n(2,2)} \binom{\theta(3)}{n(3,1)} \\*
&\quad \times \binom{s-\theta(1)-\theta(2)-\theta(3)}{n(4,1)}\bigg/\binom{s}{m}.
\end{align*}
\end{assump}
\noindent As one algorithm to implement this randomization process, the experimenter could assign each individual in the experiment a random number and then choose the $m$ lowest values to be assigned to the intervention arm. There are multiple related ways to implement the urn randomization process, such as stratification on the basis of covariates, which imply different functional forms of $f$. To complete the derivation of the specific likelihood for the urn case, I substitute the functional form of $f$ given by \ref{urn} into the general likelihood expression in (\ref{eq:f}).  Therefore, under \ref{sutva} and \ref{urn}, the likelihood of the potential outcome type configuration $\bm{\theta}$ given the data configuration $\bm{g}$, the sample size $s$, and the number randomized into the intervention arm $m$ is as follows:
\begin{align}
\mathcal{L}(\bm{\theta} | \bm{g},s,m)
= \sum_{\ell=0}^{\theta(1)}&\binom{\theta(1)}{\ell} \nonumber\\*
\times &\binom{\theta(2)}{g(2)-\ell} \nonumber\\*
\times &\binom{\theta(3)}{\theta(1)+\theta(2)+g(1)+g(2)+g(3)-s-\ell} \nonumber\\*
\times &\binom{s-\theta(1)-\theta(2)-\theta(3)}{s+\ell-\theta(1)-\theta(3)-g(2)-g(3)}\bigg/\binom{s}{m}.
\end{align}
This specific likelihood appears in \citet{copas1973}.\footnote{To translate from \citet{copas1973}, who presents this likelihood on page 469, $N \rightarrow s, s_0 \rightarrow g(1), n_0-s_0 \rightarrow g(2), s_1 \rightarrow g(3), n_1-s_1 \rightarrow s-g(1)-g(2)-g(3), n_{(1)} \rightarrow \theta(1), n_B \rightarrow \theta(2), n_A \rightarrow \theta(3),$ and $n_{AB} \rightarrow s-\theta(1)-\theta(2)-\theta(3)$. } 
However, that paper does not use the likelihood to conduct inference on the number of defiers, which is a nuisance parameter with respect to inference on other quantities.

\section{Inference: Hypothesis Testing and Confidence Intervals}\label{sec:hypothesis}

Using the likelihood of the potential outcome type configuration $\bm{\theta}$, I can test hypotheses on various quantities of interest. Consider the null hypothesis that the potential outcome type configuration $\bm{\theta}$ is in the set $H_0$.  As the test statistic, I use the likelihood ratio, which is the ratio of the maximum likelihood of the potential outcome type configurations under the null hypothesis to the maximum likelihood of all possible potential outcome type configurations.  For data configuration $\bm{g}$, define the likelihood ratio of $\bm{g}$ conditional on the sample size $s$ and other parameters that govern the randomization process $\bm{\gamma}$ as follows:
\begin{align*}
\lambda(\bm{g} | s, \bm{\gamma}) = \frac{\max_{\bm{\theta} \in H_0} \mathcal{L}(\bm{\theta} | \bm{g},s,\bm{\gamma})}{\max_{\bm{\theta}}\mathcal{L}(\bm{\theta} | \bm{g},s,\bm{\gamma})}.
\end{align*}
I solve the non-linear integer programming problems in the numerator and denominator of the likelihood ratio through a grid search over the finite parameter space.  Data configurations that are more extreme under the null hypothesis yield smaller likelihood ratios.

I reject the hypothesis when the likelihood ratio $\lambda(\bm{g}|s,\bm{\gamma})$ is below a critical value. To choose the critical value, I calculate the finite sample distribution of $\lambda(\bm{G}|s,\bm{\gamma})$ under all potential outcome type configurations in the null hypothesis.  I choose the critical value as the largest number guaranteeing that the type I error rate, the probability of falsely rejecting a true null hypothesis, is below the nominal level of the test for all potential outcome type configurations in the null hypothesis.  The rejection rule is equivalent to assigning each data configuration $\bm{g}$ a p-value equal to
\begin{align*}
\max_{\bm{\theta}\in H_0} P(\lambda(\bm{G}|s,\bm{\gamma})\leq \lambda(\bm{g}|s,\bm{\gamma}) | \bm{\theta},s,\bm{\gamma})
\end{align*}
and rejecting the test exactly when this p-value is at most the nominal level of the test $\alpha$. This rejection rule yields the same results that would be obtained under a ``worst case prior'' in a Bayesian framework.  By taking the maximum over all potential outcome type configurations in the null hypothesis, it guarantees that the type I error rate is controlled.  This approach is established in the finite sample inference literature \citep{chiba2015,rigdon2015,ding2016,li2016}.  To increase power, it is possible to construct alternative rejection rules that control different error rates, such as the average type I error across potential outcome type configurations $\bm{\theta}$  in the null hypothesis.

The test can generate confidence intervals. Suppose we want to construct a one-sided confidence interval $[L, \text{ --- } ]$ on quantity of interest $q(\bm{\theta})$ at a $(1-\alpha)\%$ level. Define the lower end of the confidence interval as the smallest real number $L$ such that the data configuration $\bm{g}$ cannot reject the following hypothesis at an $\alpha\%$ level:
\begin{align*}
&H_0: q(\bm{\theta})\leq L.
\end{align*} 
Suppose we instead want to construct a one-sided confidence interval $[ \text{ --- } ,  U]$ on quantity of interest $q(\bm{\theta})$ at a $(1-\alpha)\%$ level. Define the upper end of the confidence interval as the largest real number $U$ such that the data configuration $\bm{g}$ cannot reject the following hypothesis at an $\alpha\%$ level:
\begin{align*}
&H_0: q(\bm{\theta})\geq U.
\end{align*}
The construction of two-sided confidence intervals proceeds analogously, with tests for the upper and lower bounds at the $\alpha/2\%$ instead of $\alpha\%$ level. It is possible to compute these confidence intervals because the domain of the quantity of interest $q(\bm{\theta})$ is finite, so $q(\bm{\theta})$ can only admit finitely many values. Therefore, only finitely many hypothesis tests are required to construct the confidence interval.

\section{Examples: Clinical Trials for New Drugs}\label{sec:fda}

Through hypothetical examples, I demonstrate that the procedure that I propose can perform informative inference on the number of defiers.  To do so, I conduct inference on an analogous quantity---the number of individuals who would be killed if randomized into the intervention arm. While doing so, I demonstrate that the likelihood ratio test statistic can distinguish between data configurations that yield the same point estimate of the average intervention effect.  Tests from the finite sample inference literature that use the point estimate of the average intervention effect as the test statistic would have difficulty distinguishing between such data configurations.  I also demonstrate that the inference procedure provides p-values and confidence intervals on various quantities that capture heterogeneous intervention effects, including a second quantity for which previous inference is uninformative. Finally, I demonstrate that the inference procedure can test multiple hypotheses simultaneously.

Suppose that a data analyst for the Food and Drug Administration (FDA) receives the results from randomized clinical trials for two new drugs: Vita, which treats prostate cancer, and Mortem, which treats lung cancer. Especially since the drugs treat different types of cancer, there is no reason to compare the trial results, so the task of the analyst is to recommend approval or rejection of each drug independently. In both trials, the intervention provides access to a drug, and the outcome is survival, which is equal to one if the individual is alive and zero if the individual is dead at the end of the trial. In each trial, the sample size $s$ is $100$. Each trial follows the randomization process specified in \ref{iid} such that individuals in the intervention arm are drawn independently and identically from the sample with an  intended fraction randomized into intervention of $p=0.5$.  The data configurations for each trial appear in Table~\ref{fig:exampleFDA}.

\begin{table}[thbp!]
\captionsetup{justification=centering}
\caption{Data Configurations from Two Hypothetical Clinical Trials}
\begin{subtable}[t]{0.49\textwidth}
         \centering
         \subcaption[short for lof]{Vita}
         \includegraphics[width=\textwidth]{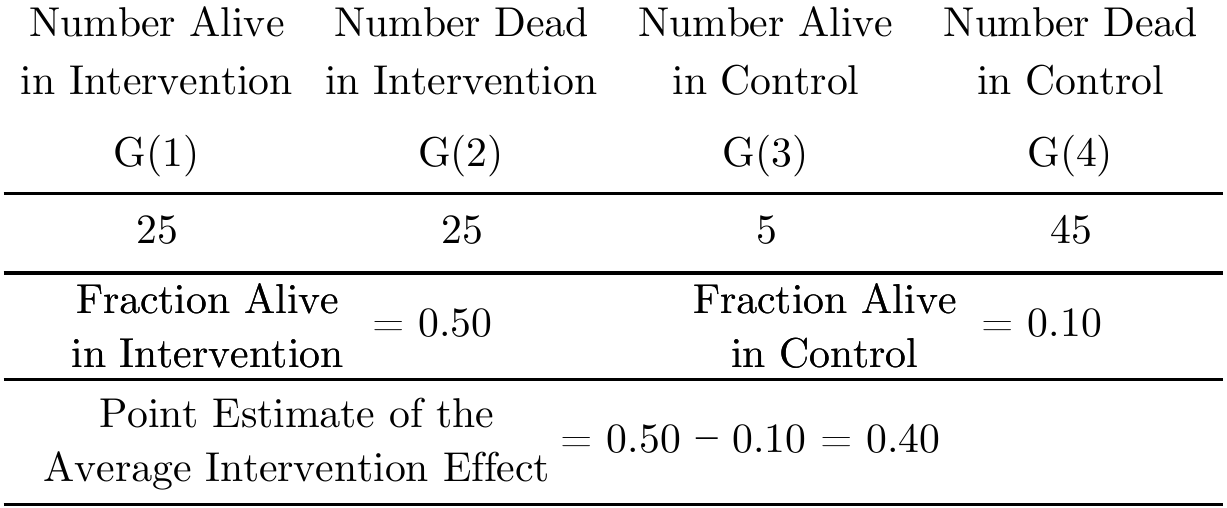}
\end{subtable}
\hfill
\begin{subtable}[t]{0.49\textwidth}
         \centering
         \subcaption[short for lof]{Mortem}
         \includegraphics[width=\textwidth]{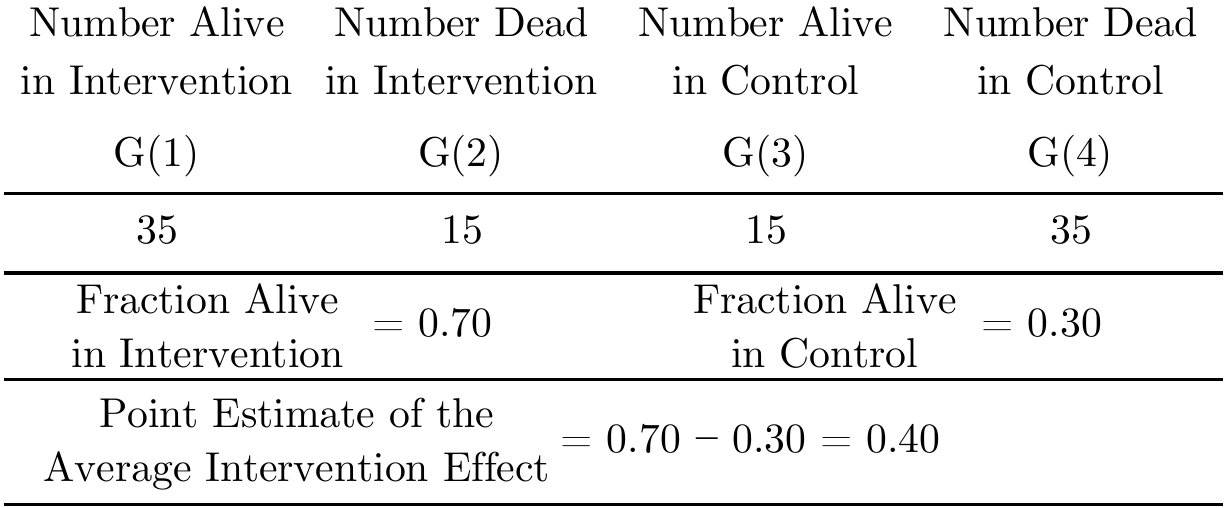}
         \end{subtable}
\label{fig:exampleFDA}
\end{table} 

The FDA could be interested in testing hypotheses about various quantities. Table \ref{fig:quantities} presents several quantities of interest, and it gives their interpretation when the outcome represents treatment, as in the model, as well as their interpretation when the outcome represents survival, as in the trials.  As shown, the number who would be killed is analogous to the number of defiers.  The average intervention effect is equal to the number who would be saved minus the number who would be killed, divided by the sample size.      

\begin{table}[hbtp!]
\captionsetup{justification=centering}
\caption{Interpretation of Quantities of Interest for Treatment and Survival Outcomes}
\begin{center}
\vspace{-5mm}
\includegraphics[width=0.9\textwidth]{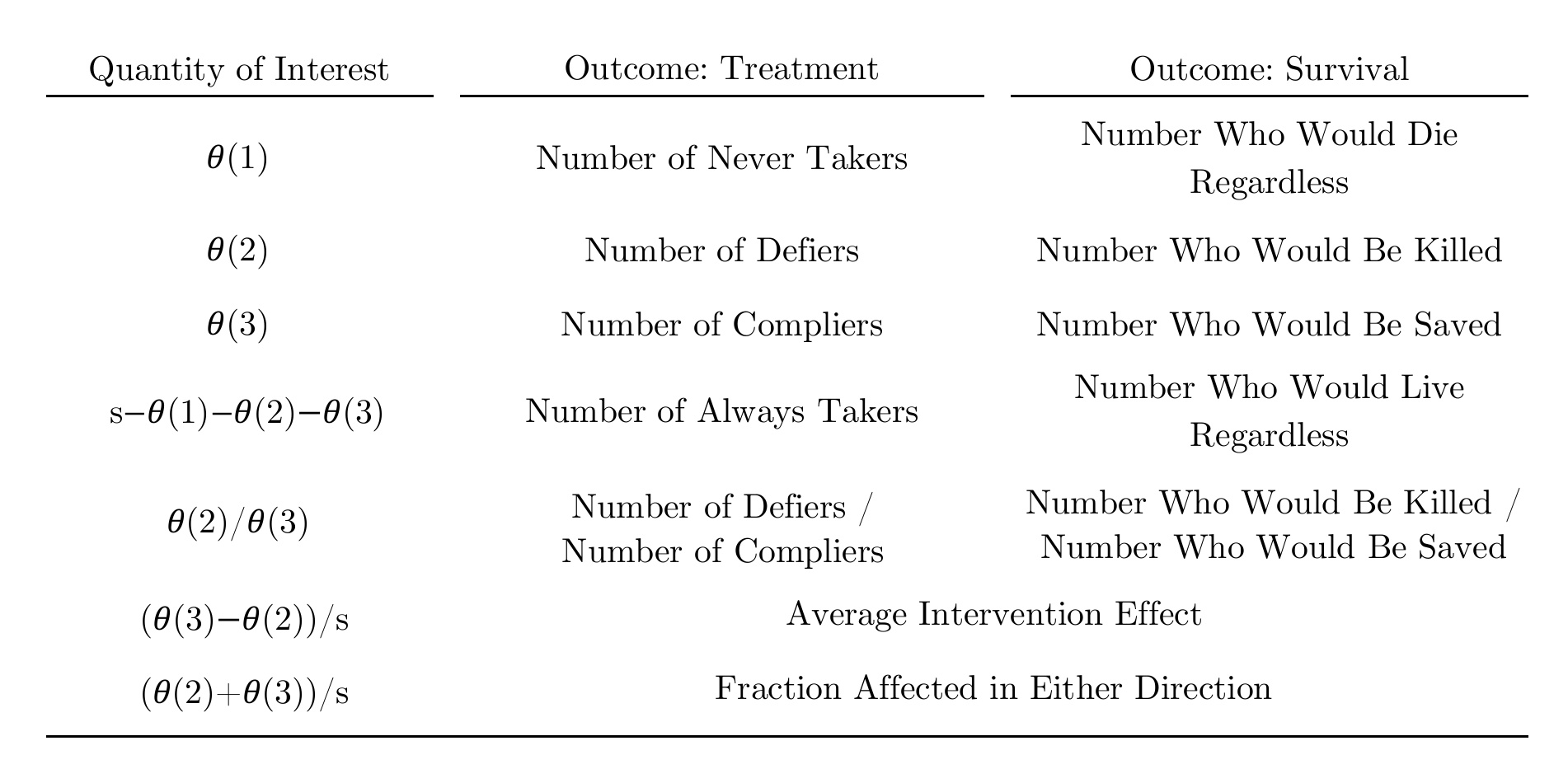}
\label{fig:quantities}
\end{center}
\vspace{-25pt}
	\begin{minipage}{0.9\linewidth}
	\scriptsize
	\emph{Note}. $s$ represents the sample size.
	\end{minipage}
\end{table}

Suppose that the FDA is only interested in efficacy, which it assesses using the average intervention effect. In both trials, the point estimate of the average intervention effect indicates a 40 percentage point increase in survival, which implies that 40 more individuals would be saved than killed in each trial.  A regression of survival on the intervention indicates that the survival increase is highly statistically significant in each trial, as shown in the first line of Table~\ref{fig:testsFDA}. This ``regression test'' relies on a simplifying asymptotic assumption, and therefore it does not necessarily control the type I error. As shown in the second line of Table~\ref{fig:testsFDA}, the proposed test of the same null hypothesis, which does not impose a simplifying asymptotic assumption and controls type I error, confirms the finding that the average intervention effect is highly statistically significant. Using either type of inference on the average intervention effect, the analyst would recommend approval of both drugs at any conventional confidence level. 

Suppose instead that the FDA is only interested in whether access to a drug would kill any individuals, which is plausible because the entire focus of phase I clinical trials is safety.  In trials of all phases, the FDA assesses safety by monitoring secondary outcomes that capture side effects, and it shuts down trials if the side effects seem too large.  Without data on secondary outcomes, previous methods cannot inform whether access to the drug would kill any individuals if the point estimate of the average intervention effect indicates an increase in survival.  However, the procedure that I propose can test the null hypothesis that no individuals would be killed even if secondary outcomes are not available.  As shown in the third line of Table~\ref{fig:testsFDA}, the data configuration from the Vita trial does not reject the hypothesis that no one would be killed at any level, whereas the data configuration from the Mortem trial rejects this hypothesis at the 2.8\% level. I cannot infer with any confidence that access to Vita would kill any individuals, but I can infer with 95\% confidence that access to Mortem would kill at least three of the 100 individuals in the sample. By analogy, this result demonstrates that is it possible to perform informative inference on the number of defiers. If the FDA's only criterion for rejection is evidence that access to the drug would kill individuals, then the analyst would recommend approval of Vita but would recommend rejection of Mortem with 97.2\% confidence.

\begin{table}[thb!]
\captionsetup{justification=centering}
\caption{Inference using Data Configurations from Hypothetical Clinical Trials}
\begin{center}
\vspace{-7mm}
\includegraphics[width=0.95\textwidth]{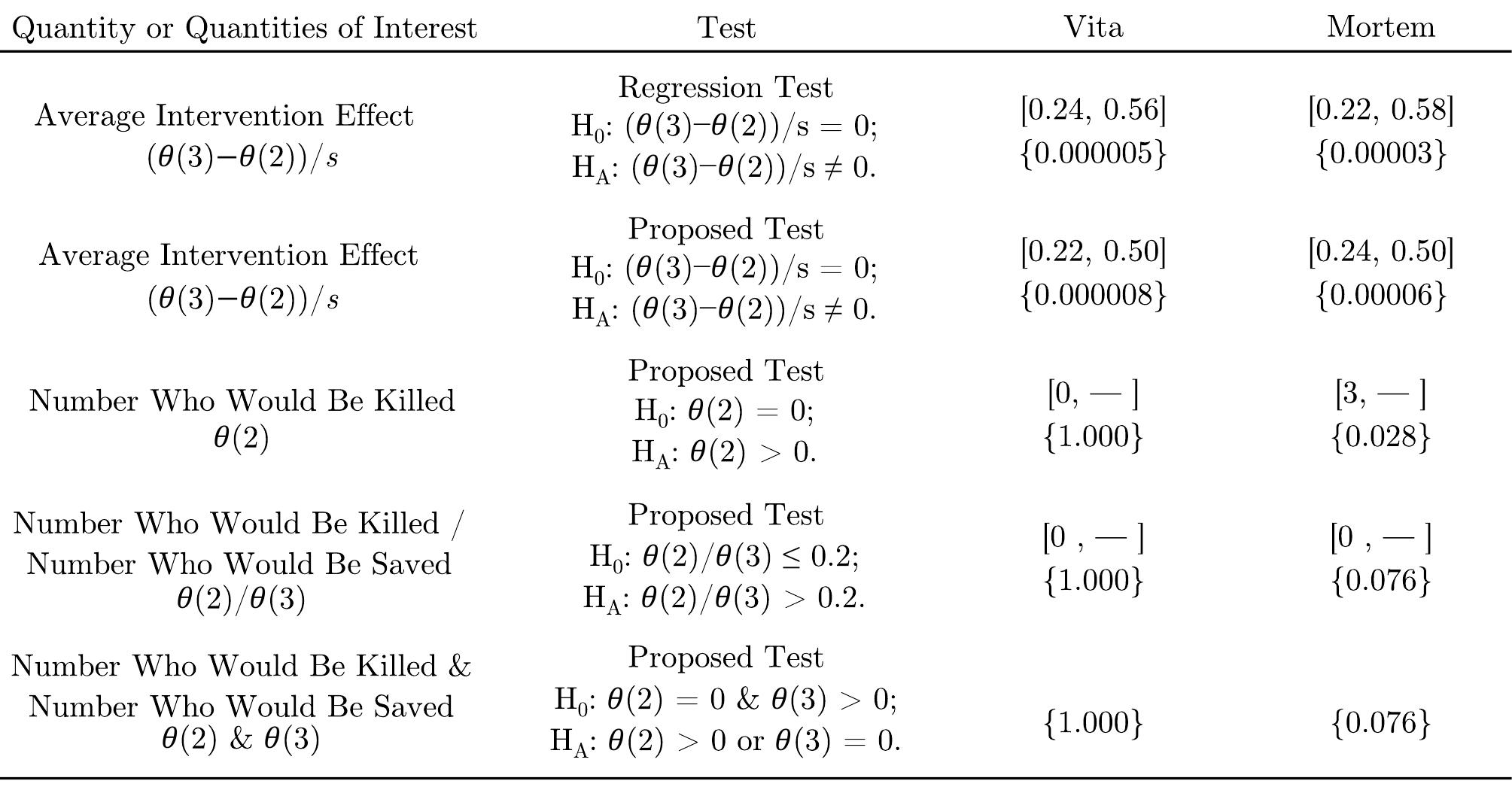}
\label{fig:testsFDA}
	\begin{minipage}{1\linewidth}
	\scriptsize
	\emph{Note}. $s$ represents the sample size. p-values for the hypothesis test are in curly braces. 95\% confidence intervals corresponding to the hypothesis test are in square brackets. A dash (---) indicates that the confidence interval is one-sided. When $\theta(2)=0$ and $\theta(3)>0$, the ratio $\theta(3)/\theta(2)$ is evaluated as infinity. When $\theta(2)=\theta(3)=0$, the ratio $\theta(3)/\theta(2)$ is ill-defined, so such potential outcome type configurations are never labeled as elements of the null hypothesis when conducting inference on $\theta(3)/\theta(2)$.  Because the joint confidence region on the number who would be killed and the number who would be saved is two-dimensional, it is not reported in the last row.
	\end{minipage}
\end{center}
\end{table}

\begin{figure}[phbt!]
	\caption{Some Potential Outcome Type Configurations Compatible with \\ the Data Configurations from the Vita and Mortem Trials}
	\captionsetup[subfigure]{labelformat=empty} 
	\vspace{-18pt} 
	\setcounter{subfigure}{0}
	\begin{center}
	\begin{subfigure}[b]{.433\linewidth}
		\centering
		\caption{Vita: No one affected}
		\includegraphics[width=1\linewidth]{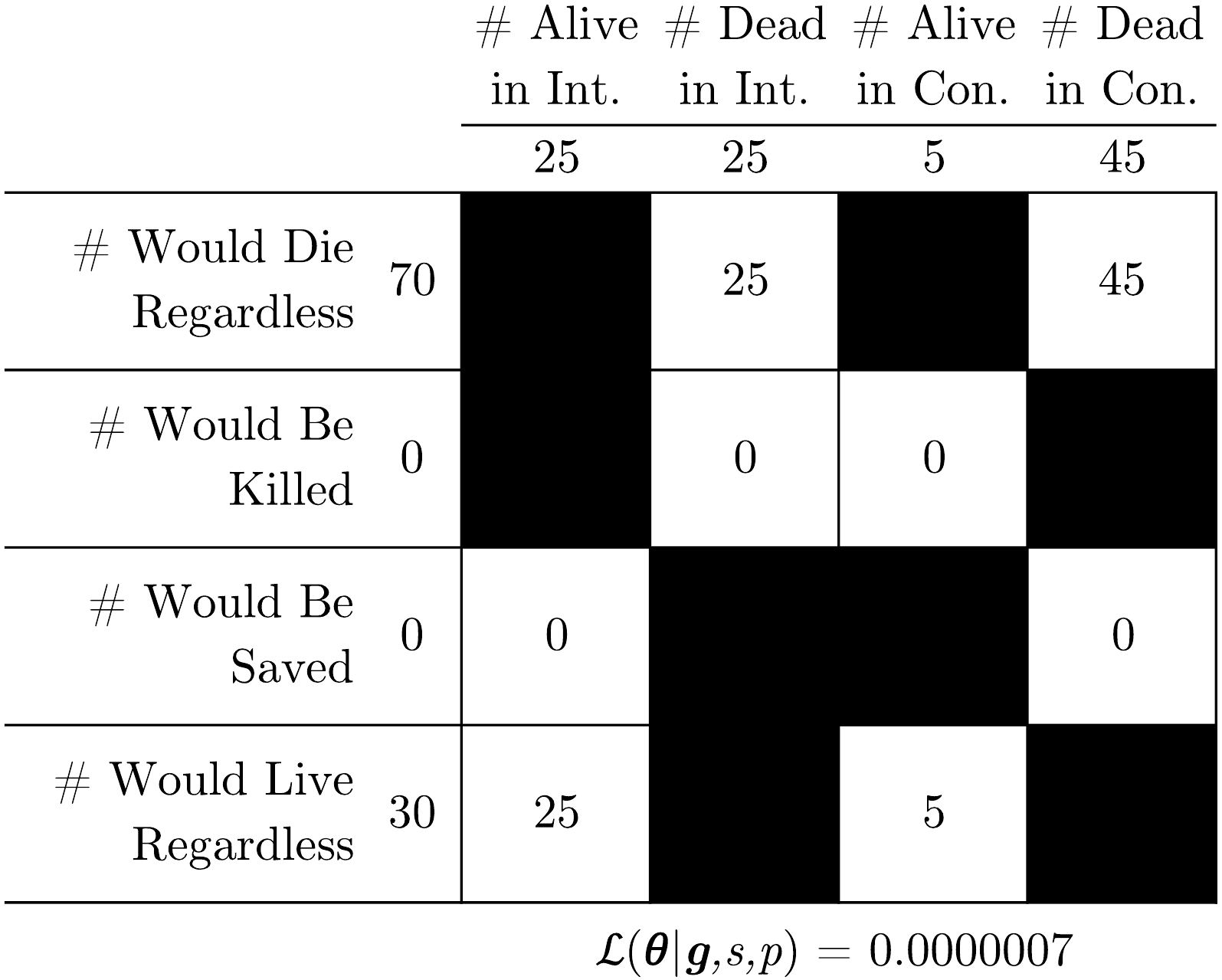}
		\label{fig:configurations.fisher.vita}
	\end{subfigure}
	\hfill
	\begin{subfigure}[b]{.433\linewidth}
		\centering
		\caption{Mortem: No one affected}		
		\includegraphics[width=1\linewidth]{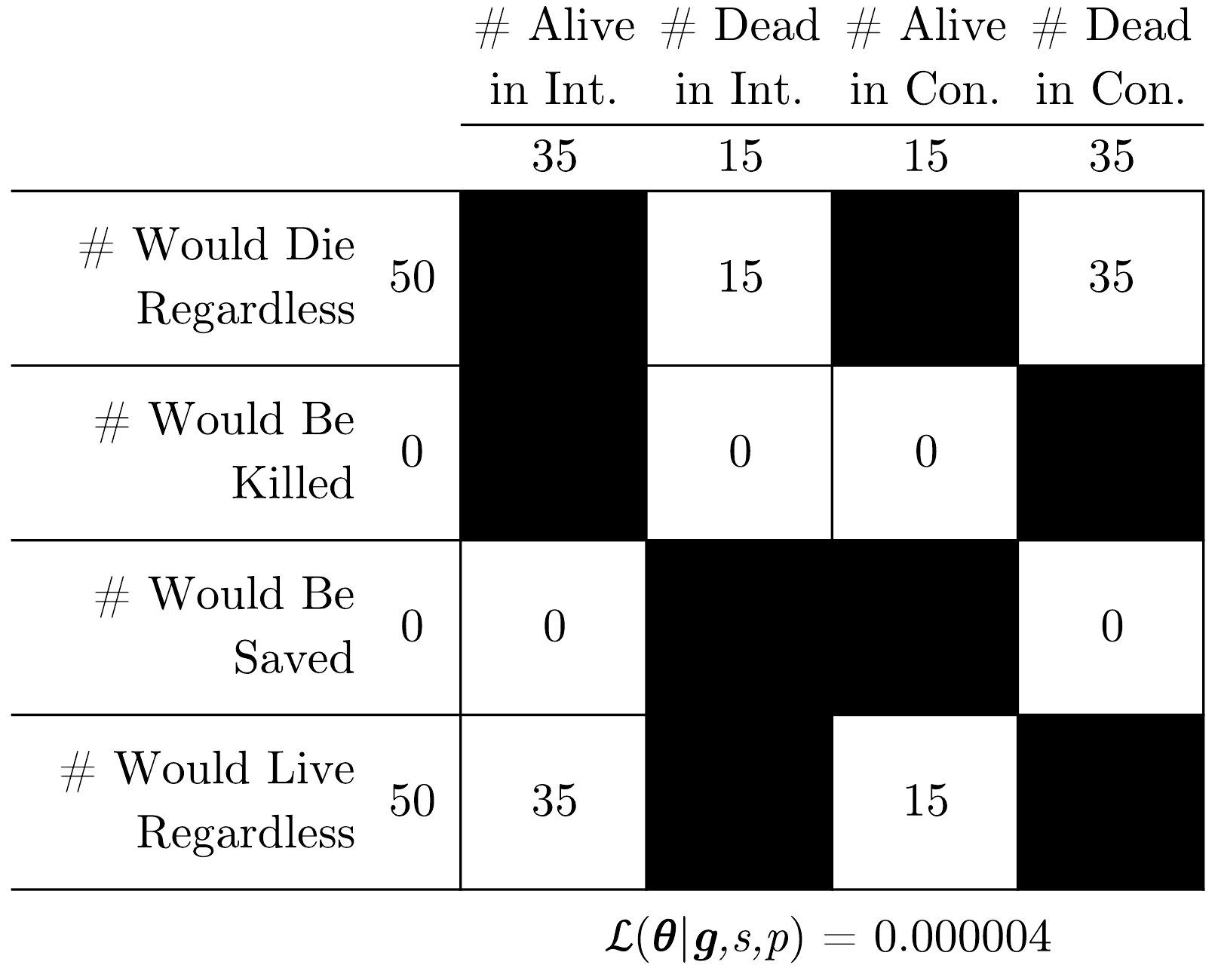}
		\label{fig:configurations.fisher}
	\end{subfigure}
	\vspace{-5pt}
	\begin{subfigure}[b]{.433\linewidth}
		\centering
		\caption{Vita: No one would be killed and \\ balanced randomization within each type}
		\includegraphics[width=1\linewidth]{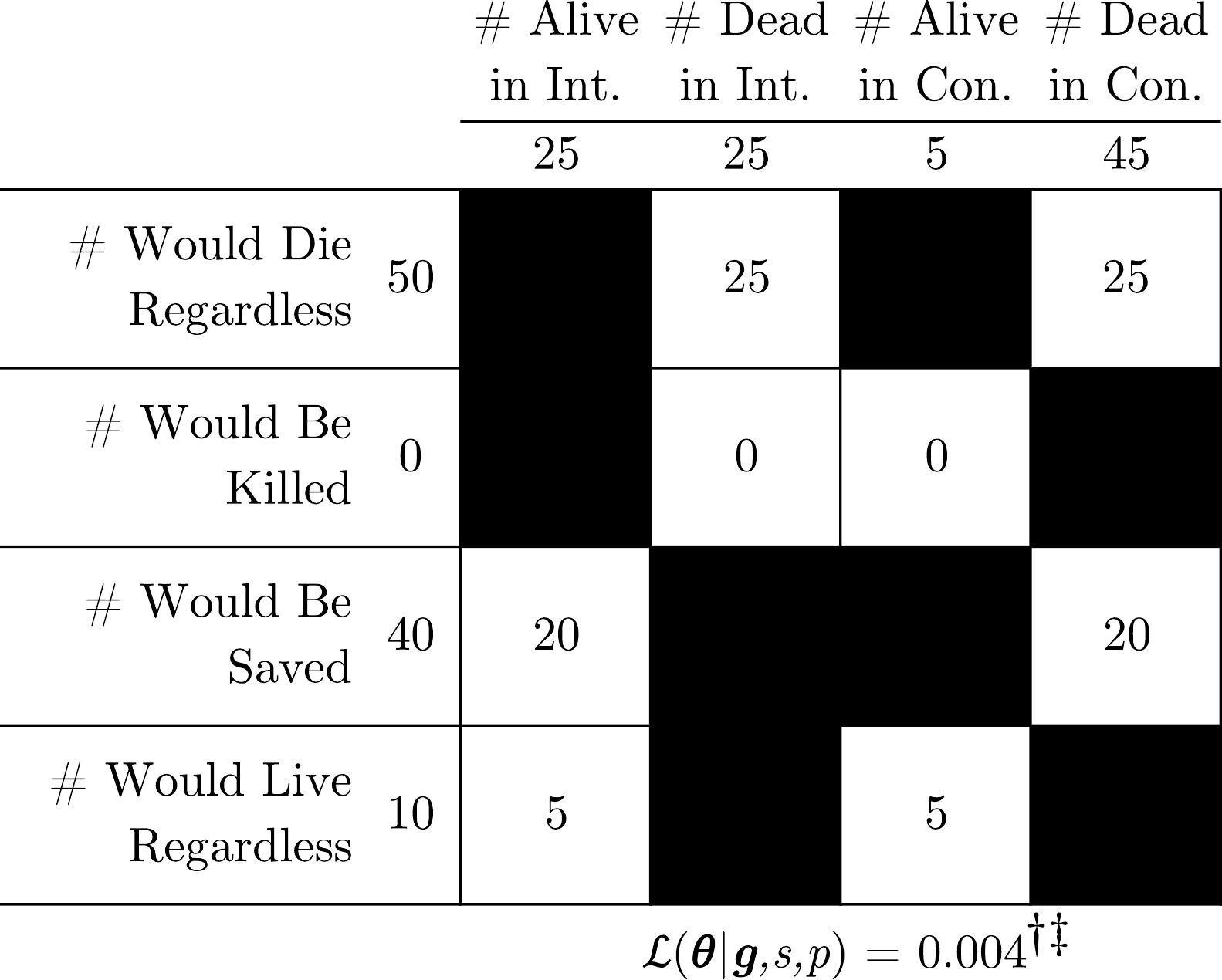}
		\label{fig:configurations.noneKilled.vita}
	\end{subfigure}
	\hfill
	\vspace{-5pt}
	\begin{subfigure}[b]{.433\linewidth}
		\centering
		\caption{Mortem: No one would be killed and \\ balanced randomization within each type}
		\includegraphics[width=1\linewidth]{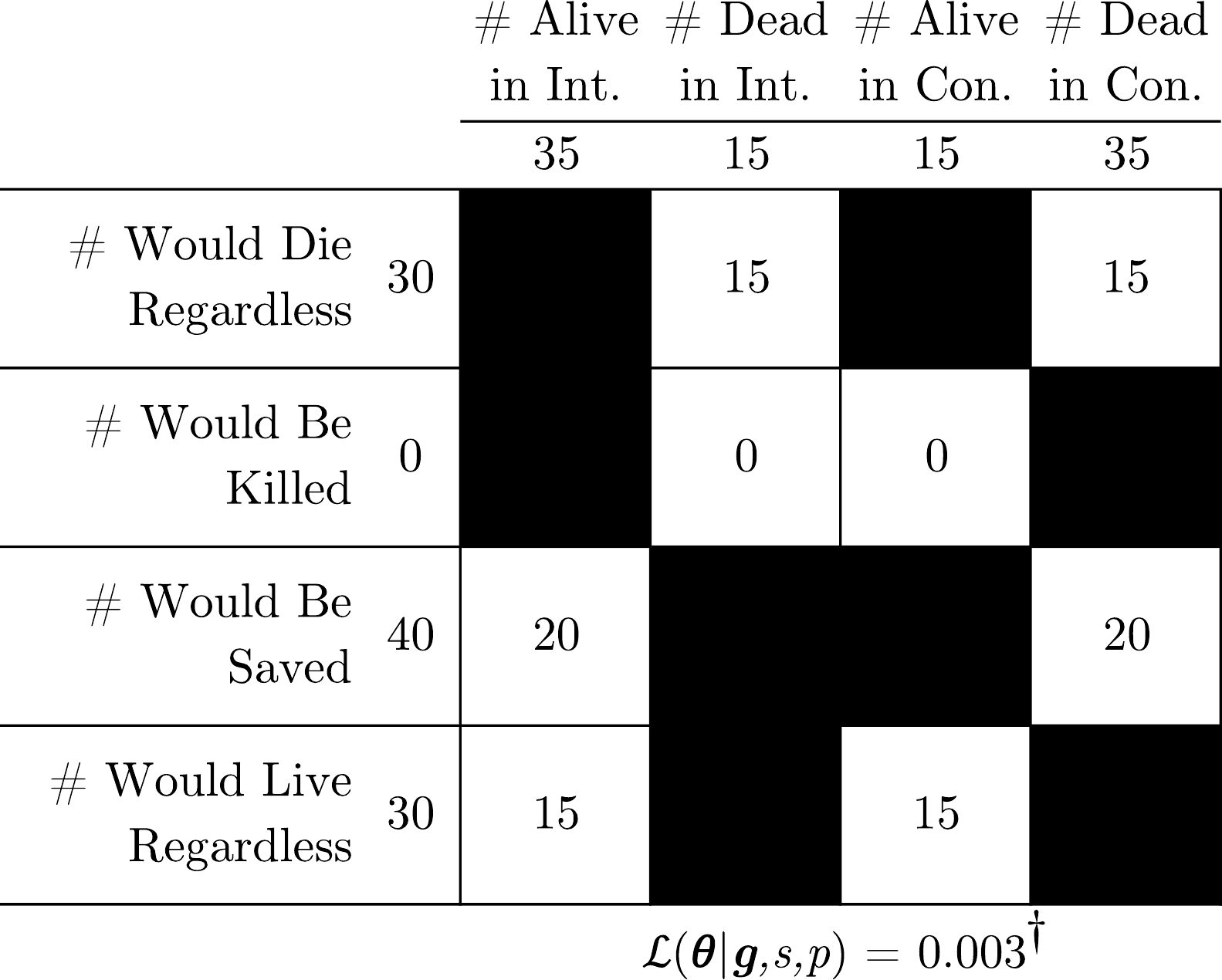}
		\label{fig:configurations.noneKilled}
	\end{subfigure}
	\begin{subfigure}[b]{.433\linewidth}
		\centering
		\caption{Vita: Everyone affected}
		\includegraphics[width=1\linewidth]{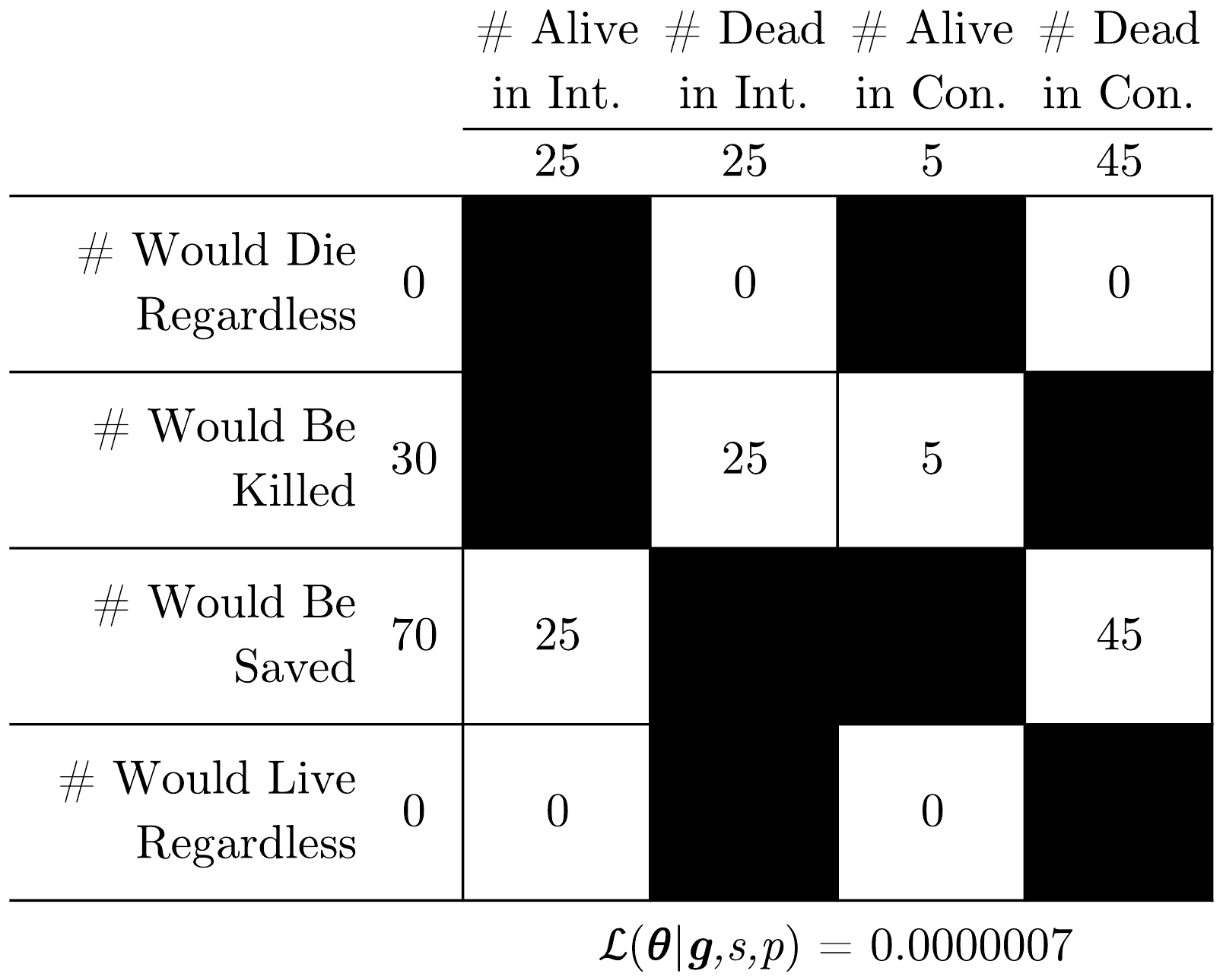}
		\label{fig:configurations.inner.vita}
	\end{subfigure}
	\hfill
	\vspace{-5pt}
	\begin{subfigure}[b]{.433\linewidth}
		\centering
		\caption{Mortem: Everyone affected}
		\includegraphics[width=1\linewidth]{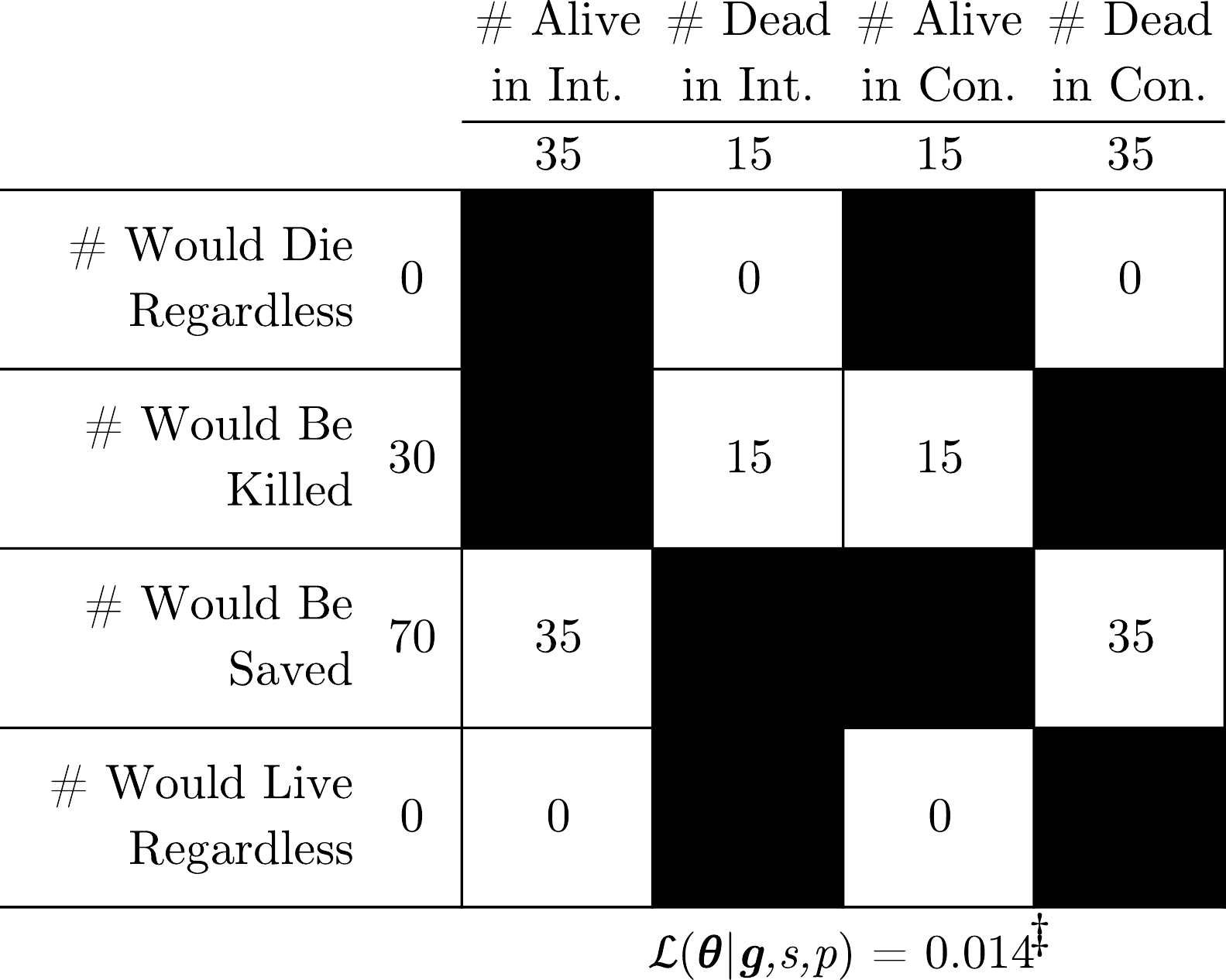}
		\label{fig:configurations.inner}
	\end{subfigure}
	\label{fig:configurations} 
	\vspace{-20pt}
	\end{center}
		\scriptsize
		\dag\ Maximum likelihood across potential outcome type configurations consistent with the null hypothesis.
		
		\ddag\ Maximum likelihood across all potential outcome type configurations.
		
		\emph{Note}.  The shaded cells equal zero. The likelihood, at the bottom of each matrix, equals the probability of the data configuration $\bm{g}$ conditional on the potential outcome type configuration $\bm{\theta}$. The sample size $s$ equals $100$, and the intended fraction randomized into intervention $p$ equals $0.5$.
\end{figure}

It is perhaps surprising that it is possible to reject the null hypothesis that no one would be killed when the point estimate of the average intervention effect indicates that there are more individuals in the trial who would be saved than killed, so I provide more detail on the intermediate inputs into the test. Figure~\ref{fig:configurations} presents three of the 45,951 potential outcome type configurations compatible with the Vita data configuration and three of the 56,151 potential outcome type configurations compatible with the Mortem data configuration.  It also presents their likelihood values.  

The first row of Figure~\ref{fig:configurations} presents the unique potential outcome type configuration in each trial that is consistent with the \citet{fisher1935} null hypothesis that no one is affected by the intervention.  Given the randomization process, we expect that half of the individuals of each potential outcome type will be randomized into the intervention arm.  However, in the potential outcome type configuration shown for Vita, 25 of 70 individuals who would die regardless are randomized into intervention (10 fewer than expected), and 25 of 30 individuals who would live regardless are randomized into intervention (10 more than expected).  Similarly, in the potential outcome type configuration shown for Mortem, 15 of 50 individuals who would die regardless are randomized into intervention (10 fewer than expected), and 35 of 50 individuals who would live regardless are randomized into intervention (10 more than expected).  The low likelihood values of approximately 0.0000007 (less than one in a million) and 0.000004 (four in a million) confirm the intuition that these potential outcome type configurations are unlikely given their respective data configurations.

The second row of Figure~\ref{fig:configurations} presents a different set of potential outcome type configurations in which no one would be killed and randomization is balanced within each potential outcome type.  Within each trial, these are the unique potential outcome type configurations that are consistent with the  ``population proportions" of each type given by \citet{imbens1997}.  The \citet{imbens1997} population proportions depend on the LATE monotonicity assumption, which implies by analogy that no one would be killed.  They also depend on a simplifying asymptotic assumption that implies balance between intervention and control within each type given the randomization process.  Given the balance, we intuit that for each trial, these potential outcome type configurations are relatively more likely than the potential outcome type configurations shown in the first row.  The likelihood values of approximately 0.004 (four in a thousand) and 0.003 (three in a thousand) confirm that these potential outcome type configurations are indeed more likely.  In fact, for both trials, these potential outcome type configurations maximize the likelihood across all configurations consistent with the null hypothesis that no one would be killed.  Therefore, the numerator of the likelihood ratio test statistic is approximately 0.004 for the Vita data configuration and approximately 0.003 for the Mortem data configuration.  For Vita, this potential outcome type configuration maximizes the likelihood across \emph{all possible} potential outcome type configurations.  Therefore, the denominator of the likelihood ratio test statistic for Vita is the same as the numerator, yielding a likelihood ratio of one.  As a result, for Vita, the likelihood ratio test cannot reject the hypothesis that no one would be killed  at any significance level.  For Mortem, however, a different potential outcome type configuration maximizes the likelihood across all possible potential outcome type configurations. 

The third row of Figure~\ref{fig:configurations} presents the unique potential outcome type configuration in each trial that is consistent with the null hypothesis that everyone is affected by the intervention.  In both trials, the potential outcome type configuration is the same.  However, this configuration implies an imbalance between intervention and control within potential outcome types in the Vita trial and a balance between intervention and control within potential outcome types in the Mortem trial.  In the Vita trial, the likelihood value of approximately 0.0000007 (less than one in a million) is smaller than the likelihood value in the second row, which is intuitive given the relative imbalance, so it is not relevant for the likelihood ratio test statistic.  However, in the Mortem trial, the likelihood value  is about 0.014 (14 in a thousand), which is the largest possible likelihood value across all potential outcome type configurations in the Mortem trial. Therefore, the denominator of the likelihood ratio for the Mortem test statistic is approximately 0.014, and the likelihood ratio test statistic for the Mortem data configuration is small and approximately equal to $0.003/0.14 = 0.21$.  This low likelihood ratio is below the critical value necessary to reject the hypothesis that no one would be killed at the 5\% level in the Mortem trial.

So far, we have supposed that the FDA is interested in efficacy \emph{or} safety, but now suppose that the FDA is interested in testing both simultaneously.  One way that the FDA might assess efficacy and safety simultaneously is to conduct inference on the ratio of the number who would be killed to the number who would be saved. To my knowledge, previous methods cannot conduct informative inference on this ratio.  Suppose the FDA is willing to approve a drug as long as no more than one individual would be killed for each five individuals who would be saved.  The trolley problem and the related transplant problem from moral philosophy \citep{foot1967, thomson1985} consider a similar criterion. To implement this criterion, the analyst could use the inference procedure that I propose to conduct inference on the null hypothesis that the ratio of the number who would be killed to the number who would be saved is less than $1/5=0.2$. Table~\ref{fig:testsFDA} shows that the data configuration of the Vita trial cannot reject the null hypothesis at any level. In contrast, the data configuration of the Mortem trial rejects the null hypothesis at a 7.6\% level. If the FDA's only criterion for rejection is evidence that access to the drug would kill no more than one individual for each five that it would save, then the analyst would recommend approval for Vita but would reject Mortem with 92.4\% confidence.

A second way that the FDA might assess efficacy and safety simultaneously is to conduct multiple hypothesis tests simultaneously. The inference procedure that I propose can do so and still control type I error, eliminating the usual concern about multiple hypothesis testing. For example, the analyst could test the composite null hypothesis that no one is killed \emph{and} at least one individual is saved. As shown in Table~\ref{fig:testsFDA}, the Vita data configuration cannot reject this hypothesis at any level, but the Mortem data configuration can reject this hypothesis at the 7.6\% level. If the FDA’s only criterion for rejection is that some individuals would be killed \emph{and} no individuals would be saved, then the analyst would recommend approval of Vita but would recommend rejection of Mortem with 92.4\% confidence.

\section{Comparisons to Related Literature: Impossibility Results} \label{sec:lit}

The inference procedure I propose is informative about the number of defiers in the sense that it can reject the null hypothesis that there are zero defiers even when the point estimate of the average intervention effect indicates that there are more compliers than defiers, but some results from the literature make such results seem impossible.  These impossibility results impose different assumptions and use different data.  In this section, I discuss how differences in assumptions and data may preclude the informative inference I demonstrate.
	
Results that apply the Boole-Fr\'echet-Hoeffding bounds \citep{boole1854, frechet1957, hoeffding1940, hoeffding1941} make informative inference on the number of defiers seem impossible.  Boole, Fr\'echet, and Hoeffding derive general bounds on the distributions of random variables, which can be applied to bound the shares of the potential outcome types in the sample.  In \ref{sec:bfhBounds}, I review how inference on the number of defiers based on the Boole-Fr\'echet-Hoeffding lower bound on the share of defiers is uninformative.  Compared to approaches that apply the Boole-Fr\'echet-Hoeffding bounds, the inference procedure I propose requires different data and assumptions.  Whereas applications of the Boole-Fr\'echet-Hoeffding bounds require only the shares treated in intervention and control, I require the four numbers of the data configuration.  Furthermore, the Boole-Fr\'echet-Hoeffding bounds can be applied under assumptions that differ from the assumptions I impose.  For example, the Boole-Fr\'echet-Hoeffding bounds can be applied under the assumption that the intervention is independent from the potential outcomes without assuming that the randomization process is known (see, for example, \citet{tian2000}).  To compare the roles of requiring different data relative to different assumptions, in \ref{sec:bfhLimit}, I derive an alternative likelihood using the data required to apply the Boole-Fr\'echet-Hoeffding bounds.  In \ref{sec:bfhEmpirical}, I demonstrate that this likelihood can still be used for informative inference on the number of defiers, suggesting that the different assumptions I require are instrumental to conducting informative inference on this quantity.

Results that impose simplifying asymptotic assumptions can also make informative inference on the number of defiers seem impossible \citep{balke1997,imbens1997,heckman2001instrumental,heckman2005,richardson2010,huber2013,huber2015testing,kitagawa2015, mourifie2017,machado2019}.  Simplifying asymptotic assumptions can be difficult to understand and critique because they are often made implicitly.  In \ref{sec:asymptoticLikelihood}, I engage with literature that requires simplifying asymptotic assumptions by adding an explicit asymptotic assumption to the model.  In \ref{sec:asymptoticEquivalence}, I show that the model plus a specific version of the asymptotic assumption is sufficient to derive likelihoods that are equivalent to likelihoods that have been derived by  \citet{barnard1947} and  \citet{kline2020}.  They do not use these likelihoods for inference on quantities for which informative inference is impossible, and they offer more straightforward derivations under different assumptions, but I start with the assumptions of the model to enhance comparability between their asymptotic likelihoods and the finite sample likelihood that I derive. I conclude in \ref{sec:asymptoticImpossibility} with impossibility results that show that these likelihoods cannot be used for informative inference on the number of defiers.

\section{Conclusion}\label{sec:conclusion}

Experiments play an important role in medicine and an increasingly important role in the social sciences, as they are often seen as the ``gold standard'' for evidence. I use experimental data to learn about heterogeneous intervention effects.  To do so, I exploit the structure of the randomization process within an experiment.  This randomization process is part of the experimental design, so it is possible to build a compelling justification for the assumption that follows from it. Even in natural experiments, it might be reasonable to exploit an assumed randomization process.

I propose a finite sample inference procedure that allows me to test hypotheses and construct confidence intervals on quantities that capture heterogeneous intervention effects. I demonstrate that I can conduct informative inference on quantities for which previous methods are uninformative, including the number of defiers---individuals whose treatment runs counter to the intervention.  I can also use the same procedure to conduct inference on other quantities, such as the average intervention effect, for which previous methods are informative but approximate or only applicable to a single quantity.  

I demonstrate in hypothetical examples that inference on a quantity analogous to the number of defiers---the number of individuals who would be killed if randomized into the intervention arm---can be useful for the drug approval process.  The presence of individuals who would be killed by an intervention that saves lives on average can pose an ethical dilemma: is it permissible to scale up an experimental intervention that kills some to save others?  The inference procedure that I propose can help to uncover the presence of such a dilemma.  In my ongoing research, I am extending the procedure with the goal of mitigating such a dilemma. By incorporating additional data on covariates, secondary outcomes, and treatment takeup, I can perform richer inference.  The main goal of such inference is to target interventions toward individuals likely to be saved and away from individuals likely to be killed.  Such inference can also provide novel tests of econometric models.  Notably, in my ongoing work, I consider a model that incorporates data on treatment takeup in which I can extend the inference procedure that I propose to test the ``exclusion restriction''  \citep{angrist1996} separately from the LATE monotonicity assumption.  

The inference procedure that I propose opens up other important areas for future work.  One is to collect data on the randomization processes for existing experiments and to analyze them using the procedure I propose. Another is to develop alternatives to asymptotic assumptions that simplify computational burdens without making informative inference on defiers impossible.

\setcounter{figure}{0}\renewcommand\thefigure{\Alph{subsection}.\arabic{figure}}
\setcounter{table}{0} \renewcommand\thetable{\Alph{subsection}.\arabic{table}}

\renewcommand\thesubsection{Appendix \Alph{subsection}}
\renewcommand\thesubsubsection{\thesubsection .\arabic{subsubsection}}

\titleformat{\section}
  {\bfseries\Large}
  {\thesection}{1em}{}

\titleformat{\subsection}
  {\bfseries\large}
  {\thesubsection}{1em}{\normalfont\large}

\titleformat{\subsubsection}
  {\bfseries}
  {\thesubsubsection}{1em}{\normalfont}

\vspace{5mm}
\section*{Appendix} \label{sec:appendix}

\addcontentsline{toc}{section}{Appendix}

\subsection{Comparison to Boole-Fr\'echet-Hoeffding Bounds}
\label{sec:bfh}

\subsubsection[Informative Inference on Defiers Applying the Boole-Fr\'echet-Hoeffding Bounds is Impossible]{Informative Inference on Defiers Applying the Boole-Fr\'echet-Hoeffding Bounds \\ is Impossible} \label{sec:bfhBounds}

I review here how applications of the Boole-Fr\'echet-Hoeffding bounds make informative inference on the number of defiers seem impossible.  To do so, I introduce additional notation because the primitive objects of the bounds are  functions of the primitives of the model.  Let $v$ represent the probability of treatment in the intervention arm $P(D=1 \text{ if } Z=1)$, which is $(s-\theta(1)-\theta(2))/s$ in the model, and let $c$ represent the probability of treatment in the control arm  $P(D=1 \text{ if } Z=0)$, which is $(s-\theta(1)-\theta(3))/s$ in the model.  I can express the Boole-Fr\'echet-Hoeffding lower bounds on the shares of defiers $\theta(2)/s$ and compliers $\theta(3)/s$ in terms of the average intervention effect, which in this notation is $v-c$:
\begin{align}
\max\Big\{ - (v - c), 0 \Big\} &\leq \theta(2)/s
 \label{eq:LDefiers} \\*
\max\Big\{ (v - c), 0 \Big\} &\leq \theta(3)/s \label{eq:LCompliers}
\end{align}

\noindent These expressions have been applied numerous times in the literature \citep{heckman1997, tian2000, zhang2003, mullahy2018, ding2019}.\footnote{To translate from \citet{heckman1997}, who give the lower bounds on compliers and defiers on page 502, $P_{E\cdot} \rightarrow v$, $P_{\cdot E} \rightarrow c$, $P_{NE} \rightarrow \theta(2)/s$, and $P_{EN} \rightarrow \theta(3)/s$.  To translate from \citet{tian2000}, who give the lower bound on the share of compliers in equation (14), $P(y_x) \rightarrow v$, $P(y_{x^\prime}) \rightarrow c$, and $PNS \rightarrow \theta(3)/s$. To translate from \citet{zhang2003}, who give the lower bound on the share of compliers following equation (8), $P_{TG} \rightarrow v$, $P_{CG} \rightarrow c$, and $\pi_{DG} \rightarrow \theta(2)/s$.  To translate from \citet{mullahy2018}, who gives the lower bound on the share of compliers in equation (26), $\pi_1 \rightarrow v$, $\pi_0 \rightarrow c$, and $\pi_{01} \rightarrow \theta(3)/s$.  To translate from \citet{ding2019}, who give the lower bound on the number of defiers in Proposition 4, $\tau \rightarrow v - c$, $N_{01} \rightarrow \theta(2)$, and $N \rightarrow s$.  \citet{manski1997mixing} and \citet{fan2010} also apply the Boole-Fr\'echet-Hoeffding bounds to conduct inference on treatment effects, but they do not supply explicit bounds on the shares of compliers and defiers.}  

These bounds can be estimated using the data configuration.  The probability of treatment in the intervention arm $v$ can be estimated with the share treated in intervention $G(1)/(G(1)+G(2))$, and the probability of treatment in the control arm $c$ can be estimated with the share treated in control $G(3)/(s-G(1)-G(2))$.  By combining these estimators, I can construct estimators for the Boole-Fr\'echet-Hoeffding lower bounds on the shares of compliers and defiers in terms of the estimator for the average intervention effect $\frac{G(1)}{G(1)+G(2)} - \frac{G(3)}{s-G(1)-G(2)}$.  An estimator for the lower bound on the share of defiers $\theta(2)/s$ is as follows:
\begin{align}
\max\Bigg\{ - \bigg( \frac{G(1)}{G(1)+G(2)} - \frac{G(3)}{s-G(1)-G(2)} \bigg) , 0 \Bigg\}.\label{eq:LDefiersEst} 
\end{align}

\noindent An estimator for the lower bound on the share of compliers $\theta(3)/s$ is as follows:
\begin{align}
\max\Bigg\{ \bigg(\frac{G(1)}{G(1)+G(2)} - \frac{G(3)}{s-G(1)-G(2)} \bigg), 0 \Bigg\}. \label{eq:LCompliersEst}
\end{align}

\noindent In my notation, the average intervention effect is equal to the share of compliers $\theta(3)/s$ minus the share of defiers $\theta(2)/s$, so a positive point estimate of the average intervention effect implies that there are more compliers than defiers.  In this case, the estimated Boole-Fr\'echet-Hoeffding lower bound on the share of defiers in (\ref{eq:LDefiersEst}) is zero, implying that the lower bound on the number of defiers is also zero.  Therefore, informative inference on the number of defiers is impossible.  Similarly, informative inference on the ratio of defiers to compliers is also impossible.  When the point estimate of the average intervention effect is negative, the lower bound on the share of compliers in (\ref{eq:LCompliersEst}) is zero, implying that informative inference on the number of compliers is impossible.

\subsubsection{Derivation of Alternative General Likelihood Using the Limited Data Required to Apply the Boole-Fr\'echet-Hoeffding Bounds}
\label{sec:bfhLimit}

While I use both more data and different assumptions than those who apply the Boole-Fr\'echet-Hoeffding bounds, I now demonstrate that the assumptions I impose are sufficient to conduct informative inference on the number of defiers, even with more restricted data.  To show this, I derive an alternative likelihood in which the available data is limited to the shares treated in the intervention and control arms, $G(1)/(G(1)+G(2))$ and $G(3)/(s-G(1)-G(2))$.  I denote realizations of these shares with $\widehat{v}$ and $\widehat{c}$.  I derive an expression for the alternative likelihood of the potential outcome type configuration $\bm{\theta}$ conditional on realizations of the shares treated in intervention and control $\widehat{v}$ and $\widehat{c}$, the sample size $s$, and other parameters that govern the randomization process from the sample into the intervention $\bm{\gamma}$ in the equations below.   

First, I express the alternative likelihood in terms of the probability distribution of the shares randomized into intervention and control in (\ref{eq:sharesG}).  Multiple data configurations can produce the same shares randomized into intervention and control, so I express the alternative likelihood as a sum of the probabilities of those data configurations indexed by $g(1)$ in (\ref{eq:sharesSum}).  By rearranging the expression, I can express the alternative likelihood as a sum over the joint distribution of the data configuration, which I do in (\ref{eq:sharesJoint}).  Finally, in (\ref{eq:sharesLike}), I replace the joint distribution of the data configuration with the likelihood expression derived in Section~\ref{sec:model} that relies on the full data configuration:
\begin{align}
\mathcal{L}(\bm{\theta} \mid \widehat{v}, \widehat{c}, s, \bm{\gamma})
&= P\bigg(\frac{G(1)}{G(1)+G(2)} = \widehat{v}, \frac{G(3)}{s-G(1)-G(2)} = \widehat{c} \mid \bm{\theta}, s, \bm{\gamma}\bigg) \label{eq:sharesG} \\
&= \sum_{g(1)=0}^{s} P\bigg(G(1)=g(1),\ \frac{G(1)}{G(1)+G(2)} = \widehat{v},\  \frac{G(3)}{s-G(1)-G(2)} = \widehat{c} \mid \bm{\theta}, s, \bm{\gamma}\bigg) \label{eq:sharesSum} \\
&= \sum_{g(1)=0}^{s} P\bigg(G(1)=g(1),\ G(2)=g(1)\Big(\frac{1}{\widehat{v}} - 1\Big),\ G(3) = \widehat{c} \Big(s-\frac{g(1)}{\widehat{v}}\Big) \mid \bm{\theta}, s, \bm{\gamma}\bigg) \label{eq:sharesJoint} \\
&= \sum_{g(1)=0}^{s} \mathcal{L} ( \bm{\theta} \mid \tilde{\bm{g}}, s, \bm{\gamma} ) \nonumber\\*
&\qquad \text{where } \tilde{\bm{g}} = \bigg(g(1),\ g(1)\Big(\frac{1}{\widehat{v}}-1\Big),\ \widehat{c}\Big(s-\frac{g(1)}{\widehat{v}}\Big), \nonumber\\*
&\qquad\qquad\qquad\quad\;\ s - g(1) - g(1)\Big(\frac{1}{\widehat{v}}-1\Big) - \widehat{c}\Big(s-\frac{g(1)}{\widehat{v}}\Big) \bigg) \label{eq:sharesLike}
\end{align}
 
\subsubsection{Informative Inference on Defiers Using the Limited Data Required to Apply the Boole-Fr\'echet-Hoeffding Bounds Is Possible but Empirically Less Precise}
\label{sec:bfhEmpirical} 

I demonstrate that the alternative likelihood in (\ref{eq:sharesLike}) can be used for informative inference on the number of defiers, and I show that inference using this alternative likelihood is empirically less precise than inference using the procedure I propose. In Table~\ref{fig:gRatio}, I compare two approaches to testing the null hypothesis that there are no defiers in the Mortem example from Section~\ref{sec:fda}.  In the first row, I reproduce the results of the test that utilizes the full data configuration, as also shown in Table~\ref{fig:testsFDA}.  In the second row, I present the results of an alternative test that utilizes the limited data of the shares treated in intervention and control, using the alternative likelihood in (\ref{eq:sharesLike}).  Both the proposed and alternative tests can conduct informative inference on the number of defiers. The p-value increases slightly when using the more limited data such that the alternative test rejects the null hypothesis at the 5.3\% level, as opposed to the 2.8\% level under the proposed test. Using more limited data obscures some, but not all, of the available information because there is still a large number of possible shares treated in intervention and control (160,131) compared to the possible number of data configurations (176,851). This exercise shows that limiting the data to the shares treated in intervention and control used in applications of the Boole-Fr\'echet-Hoeffding bounds does not preclude informative inference on the number of defiers, suggesting that the assumptions I require are instrumental to conducting informative inference.

\begin{table}[hbtp!]
\captionsetup{justification=centering}
\caption{Inference Using Full and Limited Data}
\begin{center}
\vspace{-5mm}
\includegraphics[width=0.9\textwidth]{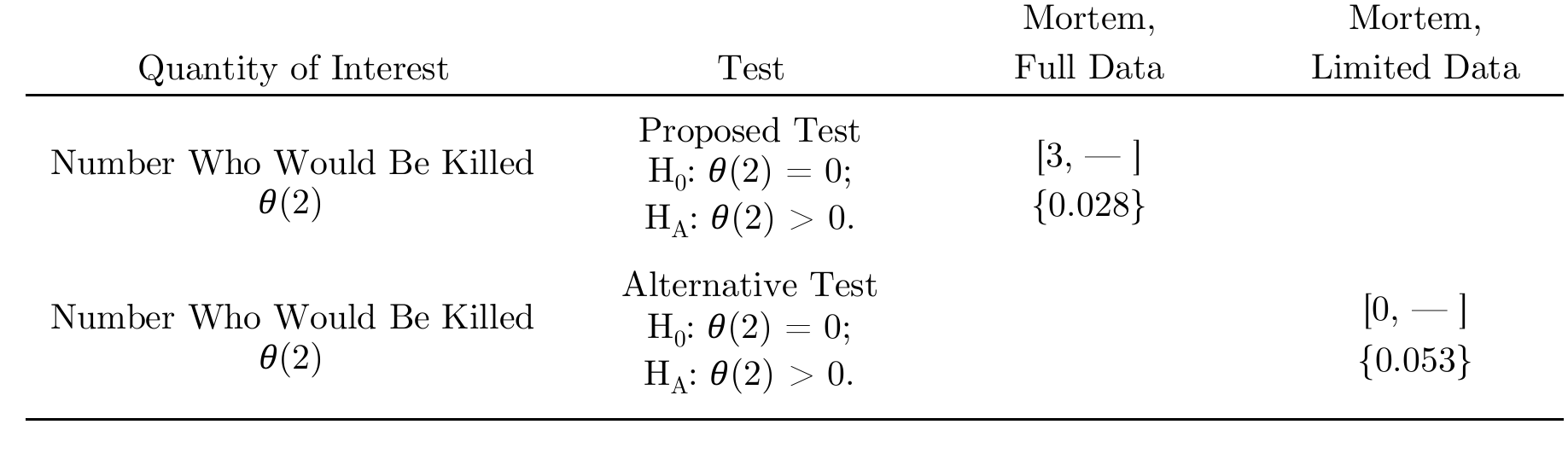}
\label{fig:gRatio}

\vspace{-10pt}
	\begin{minipage}{0.9\linewidth}
	\scriptsize
	\emph{Note}.  p-values for the hypothesis test are in curly braces. 95\% confidence intervals corresponding to the hypothesis test are in square brackets. A dash (---) indicates that the confidence interval is one-sided.  The full data consists of the four elements of the data configuration $\bm{g}$, representing the numbers of individuals alive and dead in intervention and control.  The limited data consists of the two numbers $\widehat{v}$ and $\widehat{c}$, representing the shares alive in intervention and control.
	\end{minipage}
	\end{center}
\end{table}

\subsection{Comparison to an Ancillary Simplifying Asymptotic Assumption}
\label{sec:asymptotic}

\setcounter{assump}{2}
\renewcommand{\theassump}{A.\arabic{assump}}

\subsubsection[Derivation of Alternative General Likelihood under an Ancillary Simplifying Asymptotic Assumption]{Derivation of Alternative General Likelihood under an Ancillary Simplifying \\ Asymptotic Assumption} \label{sec:asymptoticLikelihood}

Building on the two assumptions of the model, I impose an ancillary assumption that is ``asymptotic'' because it presupposes the existence of an infinite population from which the finite sample of individuals in the experiment is drawn, allowing for the possibility that the finite sample could grow.  Just as \ref{rand} specifies a randomization process \emph{within} the experiment from the finite sample, the ancillary assumption specifies a randomization process \emph{into} the experiment from an infinite population.  However, the randomization process within the experiment must occur for the experiment to be a ``randomized experiment.'' In contrast, the randomization process into the experiment need not actually occur, making an assumption based on it harder to justify.  Furthermore, even if randomization into the experiment does occur, it need not be from the infinite population required by the assumption.  The explicit ancillary asymptotic assumption that I impose for comparison to the literature is as follows:

\begin{assump} \label{samp}(Known Randomization Process into the Experiment: General Case). Individuals in the sample are selected from an infinite population through a known process, which yields a known function $h$ specifying the probability mass function of the sample potential outcome type configuration $\bm{\Theta}$ conditional on the population potential outcome type configuration $\bm{\pi}$ and the sample size $s$:
\begin{align*}
h(\bm{\theta}\mid \bm{\pi},s) \equiv P(\bm{\Theta}=\bm{\theta}\mid \bm{\pi},s). 
\end{align*}
\end{assump} 

\noindent There are many possible realizations of the sample potential outcome type configuration $\bm{\theta}$ that can be drawn from the population, so I use a capital letter to denote the random vector of the sample potential outcome type configuration $\bm{\Theta}$.  I introduce $\bm{\pi}$ to denote the \emph{population} potential outcome type configuration.  I express the elements of the population potential outcome type configuration $\bm{\pi}$ as shares instead of finite counts to allow for an infinite population.  In an infinite population, these shares remain constant as individuals are randomized into the experiment.   The assumption of no interference (\ref{sutva}) implies that individuals in the population have one of four potential outcome types.  I use $\pi(1)$ to represent the share of never takers in the population, $\pi(2)$ to represent the share of defiers in the population, $\pi(3)$ to represent the share of compliers in the population, and $1-\pi(1)-\pi(2)-\pi(3)$ to represent the share of always takers in the population. The population potential outcome type configuration $\bm{\pi}$ consists of $\pi(1), \pi(2), \pi(3),$ and $1-\pi(1)-\pi(2)-\pi(3)$.   
 
In the equations below, I derive a likelihood of the population potential outcome type configuration $\bm{\pi}$ under \ref{sutva}-\ref{samp} to yield an impossibility result. I start in (\ref{eq:piGeneral}) with a general expression for the distribution of the  data configuration $\bm{G}$ given the population potential outcome type configuration $\bm{\pi}$, the sample size $s$, and other parameters that govern the randomization process from the sample into the intervention arm $\bm{\gamma}$. Next, in (\ref{eq:piLotp}), I apply the law of total probability to represent the distribution of the data configuration as a weighted average of the distribution conditional on the sample potential outcome type configuration $\bm{\theta}$.  I then obtain an equivalent expression in terms of a likelihood of the sample potential outcome type configuration $\bm{\theta}$ in (\ref{eq:piLike}):
\begin{align}
\mathcal{L}(\bm{\pi} \mid \bm{g}, s, \bm{\gamma})
&= P \big(\bm{G} = \bm{g} \mid \bm{\pi}, s, \bm{\gamma}\big) \label{eq:piGeneral} \\
&= \sum_{\theta(1)=0}^{s}\ \sum_{\theta(2)=0}^{s-\theta(1)}\ \sum_{\theta(3)=0}^{s-\theta(1)-\theta(2)} P \big( \bm{G}=\bm{g} \mid \bm{\theta}, \bm{\pi}, s, \bm{\gamma} \big) P \big(\bm{\Theta} = \bm{\theta} \mid \bm{\pi}, s\big) \label{eq:piLotp}\\
&= \sum_{\theta(1)=0}^{s}\ \sum_{\theta(2)=0}^{s-\theta(1)}\ \sum_{\theta(3)=0}^{s-\theta(1)-\theta(2)} \mathcal{L}(\bm{\theta} \mid \bm{\pi}, \bm{g},  s, \bm{\gamma} \big) P \big(\bm{\Theta} = \bm{\theta} \mid \bm{\pi}, s\big) \label{eq:piLike}
\end{align}

\noindent   Next, I impose the assumptions of no interference (\ref{sutva}) and a known randomization process from the sample into the intervention arm (\ref{rand}).  I have shown in Section~\ref{sec:derivation} that these assumptions yield a known functional form for the likelihood of the sample potential outcome type configuration $\bm{\theta}$ conditional on the data configuration $g$, the sample size $s$, and other parameters that govern the randomization process from the sample into the intervention arm $\bm{\gamma}$.   Therefore, moving from (\ref{eq:piLike}) to (\ref{eq:piFree}), I can remove conditioning on the population potential outcome type configuration  $\bm{\pi}$ from the likelihood on the right side.  Next, in (\ref{eq:piAvg}), I impose the ancillary asymptotic assumption (\ref{samp}), which specifies a known randomization process from the infinite population into the sample, allowing me to substitute a known function  $h$ as follows: 
\begin{align}
\mathcal{L}(\bm{\pi} \mid \bm{g}, s, \bm{\gamma})
&= \sum_{\theta(1)=0}^{s}\ \sum_{\theta(2)=0}^{s-\theta(1)}\ \sum_{\theta(3)=0}^{s-\theta(1)-\theta(2)} \mathcal{L}(\bm{\theta} \mid \bm{g},  s, \bm{\gamma} \big) P \big(\bm{\Theta} = \bm{\theta} \mid \bm{\pi}, s\big) \label{eq:piFree}\\
&=  \sum_{\theta(1)=0}^{s}\ \sum_{\theta(2)=0}^{s-\theta(1)}\ \sum_{\theta(3)=0}^{s-\theta(1)-\theta(2)} \mathcal{L}(\bm{\theta} \mid \bm{g}, s, \bm{\gamma}) h(\bm{\theta} \mid \bm{\pi}, s) \label{eq:piAvg}
\end{align}

\noindent It may seem unintuitive that the imposition of an ancillary  assumption would lead to less informative inference on the number of defiers.  However, (\ref{eq:piAvg}) provides intuition by demonstrating that the ancillary assumption is a ``simplifying'' assumption.  It shows that the alternative likelihood under the ancillary assumption is a weighted average over the likelihood that I derive in Section~\ref{sec:derivation}, implying that the ancillary assumption obscures information by changing the target of inference from the sample potential outcome type configuration $\bm{\theta}$ to the population potential outcome type configuration  $\bm{\pi}$.

I substitute the general expression for the likelihood of the sample potential outcome type configuration (\ref{eq:f}) into (\ref{eq:piAvg}) to obtain a general expression for the alternative likelihood under the ancillary assumption:
\begin{align}
\mathcal{L}(\bm{\pi} \mid \bm{g}, s, \bm{\gamma})
&= \sum_{\theta(1)=0}^{s}\ \sum_{\theta(2)=0}^{s-\theta(1)}\ \sum_{\theta(3)=0}^{s-\theta(1)-\theta(2)}\ \sum_{\ell=0}^{\theta(1)} f\big(\ell,g(2)-\ell,\theta(1)+\theta(3)+g(1)+g(2)+g(3)-s-\ell, \nonumber\\*
&\qquad\qquad s+\ell-\theta(1)-\theta(3)-g(2)-g(3) \mid \bm{\theta},s,\bm{\gamma}\big) h(\bm{\theta} \mid \bm{\pi}, s). \label{eq:pifh}
\end{align}

\noindent In this equation, the imposition of specific cases of \ref{rand} and \ref{samp} implies specific functional forms for $f$ and $h$, respectively.

\subsubsection{Manipulation of Specific Alternative Likelihoods to Demonstrate Equivalence to Likelihoods from the Literature} \label{sec:asymptoticEquivalence}

I now show that specific cases of the assumptions \ref{rand} and \ref{samp} yield likelihoods from \citet{barnard1947} and \citet{kline2020}.  First, I introduce a specific randomization process from the population into the sample:

\renewcommand{\theassump}{A.3(urn)}
\begin{assump}\label{samp:urn}{(Known Randomization Process into the Experiment: Urn Case).}
Individuals in the sample are selected from an infinite population by drawing $s$ names from an urn, which implies that the sample potential outcome type configuration $\bm{\Theta}$ follows a multinomial distribution parameterized by the population potential outcome type configuration $\bm{\pi}$ and the sample size $s$:
\begin{align*}
h(\bm{\theta} \mid \bm{\pi},s) &= \frac{s!}{\theta(1)!\theta(2)!\theta(3)! \big(s-\theta(1)-\theta(2)-\theta(3)\big)!} \pi(1)^{\theta(1)} \pi(2)^{\theta(2)} \pi(3)^{\theta(3)} \nonumber\\*
&\qquad \times \big(1-\pi(1)-\pi(2)-\pi(3)\big)^{s-\theta(1)-\theta(2)-\theta(3)}.
\end{align*}
\end{assump}

\noindent Although I provide two specific cases of the randomization process within the experiment, \ref{iid} and urn \ref{urn}, I only provide one specific case of the randomization process into the experiment because it is sufficient for comparison to the literature.  The specific urn case that I provide yields a multinomial distribution because the population is infinite, but it would yield a different distribution if the population were finite.  An analogous IID case would yield an infinite sample from an infinite population, making it incompatible with the finite sample likelihood that I derive. 

To obtain an alternative likelihood equivalent to a likelihood in \citet{barnard1947}, I impose \ref{iid}, yielding an expression for $f$, and I impose \ref{samp:urn}, yielding an expression for $h$.  In (\ref{eq:urnIIDBig}), I substitute these expressions into (\ref{eq:pifh}), the general expression for the alternative likelihood under the ancillary asymptotic assumption: 
\begin{align}
\mathcal{L}(\bm{\pi} \mid \bm{g}, s, p)
&= \sum_{\theta(1)=0}^{s}\ \sum_{\theta(2)=0}^{s-\theta(1)}\ \sum_{\theta(3)=0}^{s-\theta(1)-\theta(2)}\ \sum_{\ell=0}^{\theta(1)} \mathrm{binom}\big(\ell,\theta(1),p\big) \mathrm{binom}\big(g(2)-\ell, \theta(2),p\big) \nonumber\\*
&\quad \times \mathrm{binom}\big(\theta(1)-\theta(3)+g(1)+g(2)+g(3)-s-\ell, \theta(3), p\big) \nonumber\\*
&\quad \times \mathrm{binom}\big(s+\ell-\theta(1)-\theta(3)-g(2)-g(3), s-\theta(1)-\theta(2)-\theta(3), p\big) \nonumber\\*
&\quad \times \frac{s!}{\theta(1)! \theta(2)! \theta(3)! \big(s-\theta(1)-\theta(2)-\theta(3)\big)!} \pi(1)^{\theta(1)} \pi(2)^{\theta(2)} \pi(3)^{\theta(3)} \nonumber\\*
&\quad \times \big(1-\pi(1)-\pi(2)-\pi(3)\big)^{s-\theta(1)-\theta(2)-\theta(3)} \label{eq:urnIIDBig}
\end{align}

\noindent I perform several steps to show that this likelihood is equivalent to a likelihood that appears in \citet{barnard1947}.  To begin, I switch the order of summation and adjust the summation limits accordingly:
\begin{align}
\mathcal{L}(\bm{\pi} \mid \bm{g}, s, p)
&= \sum_{\ell=0}^s \sum_{\theta(1)=\ell}^{s}\ \sum_{\theta(2)=0}^{s-\theta(1)}\ \sum_{\theta(3)=0}^{s-\theta(1)-\theta(2)}\ \mathrm{binom}\big(\ell,\theta(1),p\big) \mathrm{binom}\big(g(2)-\ell, \theta(2),p\big) \nonumber\\*
&\quad \times \mathrm{binom}\big(\theta(1)-\theta(3)+g(1)+g(2)+g(3)-s-\ell, \theta(3), p\big) \nonumber\\*
&\quad \times \mathrm{binom}\big(s+\ell-\theta(1)-\theta(3)-g(2)-g(3), s-\theta(1)-\theta(2)-\theta(3), p\big) \nonumber\\*
&\quad \times \frac{s!}{\theta(1)! \theta(2)! \theta(3)! \big(s-\theta(1)-\theta(2)-\theta(3)\big)!} \pi(1)^{\theta(1)} \pi(2)^{\theta(2)} \pi(3)^{\theta(3)} \nonumber\\*
&\quad \times \big(1-\pi(1)-\pi(2)-\pi(3)\big)^{s-\theta(1)-\theta(2)-\theta(3)}. \label{eq:urnIIDE1}
\end{align} 
Next, I reparameterize to remove $\theta(1)$, $\theta(2)$, and $\theta(3)$ from the expression. Without loss of generality, I replace $\theta(1)$ with $t(1)+\ell$, I replace $\theta(2)$ with $t(2)+g(2)-\ell$, and I replace $\theta(3)$ with $t(3)-t(1)-g(1)-g(2)-g(3)+s$. This reparameterization yields the following expression for the likelihood: 
\begin{align}
\mathcal{L}(\bm{\pi} \mid \bm{g}, s, p)
&= \sum_{\ell=0}^s \sum_{t(1)=0}^{s-\ell}\ \sum_{t(2)=\ell-g(2)}^{s-t(1)-g(2)}\ \sum_{\substack{t(3)=t(1)+g(1)\\+g(2)+g(3)-s}}^{g(1)+g(3)-t(2)}\ \mathrm{binom}\big(\ell,\ell+t(1),p\big) \nonumber\\*
&\quad \times \mathrm{binom}\big(g(2)-\ell, g(2)-\ell+t(2),p\big) \nonumber\\*
&\quad \times \mathrm{binom}\big(t(3), t(3)+s-g(1)-g(2)-g(3)-t(1), p\big) \nonumber\\*
&\quad \times \mathrm{binom}\big(g(1)-t(3), g(1)-t(3)+g(3)-t(2), p\big) s! \Big[\big(t(1)+\ell\big)! \big(t(2)+g(2)-\ell\big)! \nonumber\\*
&\quad \times \big(t(3)-t(1)-g(1)-g(2)-g(3)+s\big)! \big(g(1)+g(3)-t(1)-t(3)\big)! \Big]^{-1} \pi(1)^{t(1)+\ell}  \nonumber\\*
&\quad \times \pi(2)^{t(2)+g(2)-\ell} \pi(3)^{t(3)-t(1)-g(1)-g(2)-g(3)+s} \nonumber\\*
&\quad \times \big(1-\pi(1)-\pi(2)-\pi(3)\big)^{g(1)+g(3)-t(1)-t(3)}. \label{eq:urnIIDE2}
\end{align} 
Next, I change the limits of summation to reflect the fact that the summand equals zero if the first argument of any of the binomial distribution functions is negative or if the second argument of any of the binomial distribution functions is less than the first. The binomial probability mass function is still well-defined in such cases because the intended fraction randomized into intervention $p$ is constrained to be strictly between zero and one. Starting with the outer summation, we know that $\ell$ is at most $g(2)$ such that the second binomial distribution function is non-zero. Since $s\geq g(2)$, we can replace the upper limit of the outer summation with $g(2)$. Next, we know that $t(1)$ is at most $s-g(1)-g(2)-g(3)$ such that the third binomial distribution is non-zero. Since $s-g(1)-g(2)-g(3)\leq s-\ell$ for all $\ell$ between zero and $g(2)$, we can replace the upper limit of the second summation with $s-g(1)-g(2)-g(3)$. Through similar arguments, we can replace the limits in the remaining two summations: 
\begin{align}
\mathcal{L}(\bm{\pi} \mid \bm{g}, s, p)
&= \sum_{\ell=0}^{g(2)} \sum_{t(1)=0}^{\substack{s-g(1)-g(2)\\-g(3)}}\ \sum_{t(2)=0}^{g(3)}\ \sum_{t(3)=0}^{g(1)}\ \mathrm{binom}\big(\ell,\ell+t(1),p\big) \nonumber\\*
&\quad \times \mathrm{binom}\big(g(2)-\ell, g(2)-\ell+t(2),p\big) \nonumber\\*
&\quad \times \mathrm{binom}\big(t(3), t(3)+s-g(1)-g(2)-g(3)-t(1), p\big) \nonumber\\*
&\quad \times \mathrm{binom}\big(g(1)-t(3), g(1)-t(3)+g(3)-t(2), p\big) s! \Big[\big(t(1)+\ell\big)! \big(t(2)+g(2)-\ell\big)! \nonumber\\*
&\quad \times \big(t(3)-t(1)-g(1)-g(2)-g(3)+s\big)! \big(g(1)+g(3)-t(1)-t(3)\big)! \Big]^{-1} \pi(1)^{t(1)+\ell}  \nonumber\\*
&\quad \times \pi(2)^{t(2)+g(2)-\ell} \pi(3)^{t(3)-t(1)-g(1)-g(2)-g(3)+s} \nonumber\\*
&\quad \times \big(1-\pi(1)-\pi(2)-\pi(3)\big)^{g(1)+g(3)-t(1)-t(3)}. \label{eq:urnIIDE3}
\end{align} 
With these changes of indices, the summations in (\ref{eq:urnIIDE3}) are independent. Therefore, I gather terms and expand using the definition of the binomial distribution as follows:
\begin{align}
\mathcal{L}(\bm{\pi} \mid \bm{g}, s, p)
&= s! p^{g(1)+g(2)} (1-p)^{s-g(1)-g(2)} \Bigg[\sum_{\ell=0}^{g(2)} \frac{ \pi(1)^{\ell} \pi(2)^{g(2)-\ell}}{\ell! \big(g(2)-\ell\big)!} \Bigg] \nonumber\\*
&\quad \times \Bigg[ \sum_{t(1)=0}^{\substack{s-g(1)-g(2)\\-g(3)}} \frac{\pi(1)^{t(1)} \pi(3)^{s-g(1)-g(2)-g(3)-t(1)}}{t(1)! \big(s-g(1)-g(2)-g(3)-t(1)\big)!} \Bigg] \nonumber\\*
&\quad \times \Bigg[ \sum_{t(2)=0}^{g(3)} \frac{\pi(2)^{t(2)} \big(1-\pi(1)-\pi(2)-\pi(3)\big)^{g(3)-t(2)}}{t(2)! \big(g(3)-t(2)\big)!}  \Bigg] \nonumber\\*
&\quad \times \Bigg[ \sum_{t(3)=0}^{g(1)} \frac{\pi(3)^{t(3)} \big(1-\pi(1)-\pi(2)-\pi(3)\big)^{g(1)-t(3)}}{t(3)! \big(g(1)-t(3)\big)!} \Bigg] \label{eq:urnIIDFour}
\end{align}

\noindent By the binomial theorem, I can express (\ref{eq:urnIIDFour}) as follows:
\begin{align}
\mathcal{L}(\bm{\pi} \mid \bm{g}, s, p)
&= \frac{s!}{g(1)!g(2)!g(3)!(s-g(1)-g(2)-g(3))!} p^{g(1)+g(2)} (1-p)^{s-g(1)-g(2)} \nonumber\\*
&\qquad\times \big[\pi(3)+(1-\pi(1)-\pi(2)-\pi(3))\big]^{g(1)} \big[\pi(1)+\pi(2)\big]^{g(2)} \nonumber\\* 
&\qquad \times \big[\pi(2) + (1-\pi(1)-\pi(2)-\pi(3))\big]^{g(3)} \big[\pi(1)+\pi(3)\big]^{s-g(1)-g(2)-g(3)} \nonumber\\
&= \frac{s!}{g(1)!g(2)!g(3)!(s-g(1)-g(2)-g(3))!} p^{g(1)+g(2)} (1-p)^{s-g(1)-g(2)} \nonumber\\*
&\qquad\times \big[1-\pi(1)-\pi(2)\big]^{g(1)} \big[\pi(1)+\pi(2)\big]^{g(2)} \nonumber\\* 
&\qquad \times \big[1-\pi(1)-\pi(3)\big]^{g(3)} \big[\pi(1)+\pi(3)\big]^{s-g(1)-g(2)-g(3)}. \label{eq:urnIID}
\end{align}

\noindent The expression in (\ref{eq:urnIID}) is equivalent to a  likelihood from \citet{barnard1947} under a reparameterization.\footnote{To translate from \citet{barnard1947}, who presents this likelihood in equation (4), $N \rightarrow s$, $a \rightarrow g(1)$, $b \rightarrow g(2)$, $c \rightarrow g(3)$, $d \rightarrow s-g(1)-g(2)-g(3)$, $p_{a1} \rightarrow p(1-\pi(1)-\pi(2))$, $p_{a2}\rightarrow p(\pi(1)+\pi(2))$, $p_{b1} \rightarrow (1-p)(1-\pi(1)-\pi(3))$, and $p_{b2} \rightarrow (1-p)(\pi(1)+\pi(3))$.}  

Next, to obtain an alternative likelihood that is equivalent to a different likelihood from \citet{barnard1947} and to a likelihood from \citet{kline2020} under reparameterizations, I impose \ref{urn}, yielding an expression for $f$, and I impose \ref{samp:urn}, yielding an expression for $h$.  I substitute these expressions into the general likelihood (\ref{eq:pifh}) in (\ref{eq:urnurnBig}): 
\begin{align}
\mathcal{L}(\bm{\pi} \mid \bm{g}, s, m)
&= \sum_{\theta(1)=0}^{s}\ \sum_{\theta(2)=0}^{s-\theta(1)}\ \sum_{\theta(3)=0}^{s-\theta(1)-\theta(2)}\ \sum_{\ell=0}^{\theta(1)} \binom{\theta(2)}{g(2)-\ell} \binom{\theta(3)}{\theta(1)+\theta(2)+g(1)+g(2)+g(3)-s-\ell} \nonumber\\*
&\quad \times \binom{s-\theta(1)-\theta(2)-\theta(3)}{s+\ell-\theta(1)-\theta(3)-g(2)-g(3)} \frac{s!}{\theta(1)! \theta(2)! \theta(3)! \big(s-\theta(1)-\theta(2)-\theta(3)\big)!} \nonumber\\*
&\quad \times \pi(1)^{\theta(1)} \pi(2)^{\theta(2)} \pi(3)^{\theta(3)} \big(1-\pi(1)-\pi(2)-\pi(3)\big)^{s-\theta(1)-\theta(2)-\theta(3)}\bigg/\binom{s}{m} \label{eq:urnurnBig}
\end{align}
Simplification reveals that the likelihood in (\ref{eq:urnurnBig}) is equal to the likelihood that imposes \ref{iid} instead of \ref{urn}  in (\ref{eq:urnIIDBig}), divided by $p^{g(1)+g(2)}(1-p)^{s-g(1)-g(2)}\binom{s}{m}$. Therefore, because the likelihood in (\ref{eq:urnIIDBig}) is equal to the likelihood in (\ref{eq:urnIID}) as shown above, the likelihood in (\ref{eq:urnurnBig}) is equal to the likelihood in (\ref{eq:urnIID}) divided by $p^{g(1)+g(2)}(1-p)^{s-g(1)-g(2)}\binom{s}{m}$. With further simplification, I can express the resulting likelihood as follows:
\begin{align}
\mathcal{L}(\bm{\pi} \mid \bm{g}, s, m)
&= \binom{m}{g(1)} [1-\pi(1)-\pi(2)]^{g(1)} [\pi(1)+\pi(2)]^{m-g(1)} \nonumber\\*
&\qquad \times \binom{s-m}{g(3)} [1-\pi(1)-\pi(3)]^{g(3)} [\pi(1)+\pi(3)]^{s-m-g(3)} \nonumber \\*
&=  \mathrm{binom}\big(g(1),m,1-\pi(1)-\pi(2)\big) \mathrm{binom}\big(g(3),s-m,1-\pi(1)-\pi(3)\big), \label{eq:urnurn}
\end{align}
where $\mathrm{binom}(a,b,p)$ is the binomial probability mass function:
\begin{align*}
\mathrm{binom}(a,b,p) = P(A=a\mid b,p) = \binom{b}{a} p^a (1-p)^{b-a}.
\end{align*}  
The likelihood in (\ref{eq:urnurn}) is equivalent to a likelihood from \citet{barnard1947} and to a likelihood from \citet{kline2020} under reparameterizations.\footnote{To translate from \citet{barnard1947}, who presents this likelihood in equation (2), $n \rightarrow s-m$, $a \rightarrow g(1)$, $b \rightarrow g(3)$, $c \rightarrow m-g(1)$, $d \rightarrow s-m-g(3)$, $p_a \rightarrow 1-\pi(1)-\pi(2)$, and $p_b \rightarrow 1-\pi(1)-\pi(3)$.  To translate from \citet{kline2020}, who present this likelihood in equation (1), $L_w \rightarrow m$, $L_b \rightarrow s-m$, $c_w \rightarrow g(1)$, $c_b \rightarrow g(3)$, $p_{jw} \rightarrow 1-\pi(1)-\pi(2)$, and $p_{jb} \rightarrow 1-\pi(1)-\pi(3)$.} 

\subsubsection{Informative Inference on Defiers Using Specific Alternative Likelihoods under an Ancillary Simplifying Asymptotic Assumption is Impossible} \label{sec:asymptoticImpossibility}

I conclude by introducing impossibility results that show that the likelihoods in (\ref{eq:urnIID}) and (\ref{eq:urnurn}) cannot be used for informative inference on the number of defiers.   That is, if the point estimate of the average intervention effect indicates there are more compliers than defiers, inference based on the likelihoods in (\ref{eq:urnIID}) and (\ref{eq:urnurn}) cannot reject the null hypothesis of zero defiers. To do so, I will show that the set of population potential outcome type configurations that maximizes the likelihood in (\ref{eq:urnIID}) or (\ref{eq:urnurn}) always includes a population potential outcome type configuration with zero defiers when the point estimate of the average intervention effect indicates there are more compliers than defiers.  As proof, suppose $\widehat{\bm{\pi}} = (\widehat{\pi}(1), \widehat{\pi}(2), \widehat{\pi}(3), 1-\widehat{\pi}(1) - \widehat{\pi}(2) - \widehat{\pi}(3))$ is an estimate of the population potential outcome type configuration that maximizes the likelihood in (\ref{eq:urnIID}) or (\ref{eq:urnurn}).  The implied average intervention effect $\widehat{\pi}(3) - \widehat{\pi}(2)$  will equal its point estimate $g(1)/(g(1)+g(2)) - g(3)/(s-g(1)-g(2))$, which can be confirmed by maximizing the likelihood functions in (\ref{eq:urnIID}) and (\ref{eq:urnurn}) analytically. If the point estimate of the average intervention effect is positive, it indicates that there are more compliers than defiers. In this case, there is a population potential outcome type configuration that maximizes the likelihood and has no defiers, $(\widehat{\pi}(1)+\widehat{\pi}(2), 0, \widehat{\pi}(3)-\widehat{\pi}(2), 1-\widehat{\pi}(1)-\widehat{\pi}(3))$, since:
\begin{align}
\mathcal{L}(\widehat{\bm{\pi}}, s, \bm{\gamma})
& = \mathcal{L}\big(\widehat{\pi}(1) + \widehat{\pi}(2),\ 0,\ \widehat{\pi}(3)-\widehat{\pi}(2),\ 1-\widehat{\pi}(1)-\widehat{\pi}(3) \mid \bm{g}, s, \bm{\gamma}\big).\label{eq:piMax}
\end{align}
It is straightforward to verify this claim by evaluating the likelihoods (\ref{eq:urnIID}) and (\ref{eq:urnurn}) at these population potential outcome type configurations and showing that (\ref{eq:piMax}) holds.  Thus, the set of population potential outcome type configurations that maximize the likelihood includes a population potential outcome type configuration with no defiers when the point estimate of the average intervention effect indicates there are more compliers than defiers.   Therefore, informative inference on the share of defiers is impossible using the likelihoods in (\ref{eq:urnIID}) and (\ref{eq:urnurn}).   Equation (\ref{eq:piMax}) also demonstrates that informative inference on the ratio of defiers to compliers is impossible using these likelihoods.   When the point estimate of the average intervention effect is negative, a similar procedure demonstrates that informative inference on the number of compliers is not possible using these likelihoods.  These impossibility results hold for any finite sample size $s$. 

\newpage
\singlespacing
\bibliographystyle{chicago}
\bibliography{infer}

\end{document}